\documentclass{aa}
\usepackage[colorlinks=true,linkcolor=blue,citecolor=blue]{hyperref}
\usepackage{graphicx}
\usepackage{txfonts}
\usepackage{xcolor}
\usepackage{physics}
\usepackage[version=4]{mhchem}
\usepackage{subcaption}         
\usepackage{lscape}             
\usepackage{placeins}           

\usepackage{natbib}   
\usepackage{multirow}

\newcommand{\mpl}{M_{\mathrm{p}}}
\newcommand{\rpl}{R_{\mathrm{p}}}
\newcommand{\rj}{\mathrm{R_J}}
\newcommand{\mj}{\mathrm{M_J}}

\newcommand{\orcid}[1]{\unskip\,\href{https://orcid.org/#1}{\includegraphics[width=8pt]{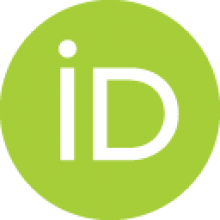}}}

\begin{document}

   \title{Thermodynamics of gas giants accreting via shocks}

   \subtitle{The role of molecular hydrogen dissociation}

   \author{Bingjie Zha\inst{1}
        \and Zhuo Chen\orcid{0000-0001-7420-9606}\inst{2}
        }

   \institute{Department of Earth Sciences, Freie Universität Berlin, Berlin 12249, Germany
            \and Chen-Ning Yang Institute for Advanced Study, Tsinghua University, Beijing 100084, P. R. China\\
            \email{chenzhuo\_astro@tsinghua.edu.cn}
            }

 
  \abstract
   {Within the core-accretion framework, gas giants acquire the bulk of their mass through accretion shocks during runaway gas accretion. The thermodynamics of these shocks—governed by the radiative cooling efficiency, accretion history, and equation of state (EoS)—sets the post-shock gas entropy and the planet's initial thermal state, strongly influencing both the observables of forming planets and their subsequent evolution.}
   {We aim to construct a coupled evolutionary model capable of predicting the initial thermodynamic conditions, intrinsic luminosities, and upper limits on the magnetic field strengths of gas giants across diverse accretion histories.} 
   {We couple radiation hydrodynamics and \ce{H2} dissociation to long-term planetary interior evolution. Specifically, we train a fast physics-informed neural network (PINN) surrogate model on 1000 first-principles simulations that resolve the radiation-hydrodynamic and thermodynamic shock processes. This surrogate model supplies the post-shock thermodynamic states as the outer boundary condition for {\tt MESA}, enabling the evolution of the planetary envelope beyond the runaway accretion phase.}
   {The surrogate model yields higher post-shock temperatures than the canonical cold-start model, with the discrepancy widening progressively with increasing planetary mass. Furthermore, the post-shock entropy exhibits a sharp jump when \ce{H2} dissociates at $\sim 2000\,$K. Gas giants evolved with the PINN-derived outer boundary conditions attain intrinsic luminosities 4--10 times higher than those predicted by cold-start models. Under optimistic assumptions, the maximum magnetic field strength at $2\,\rj$ reaches a few gauss for a $0.3\,\mj$ planet, $\sim 100\,$G for $1\,\mj$, and $\sim 500\,$G for $3\,\mj$.}
   {Incorporating detailed radiation hydrodynamics and realistic thermodynamics into planetary accretion shock models yields higher initial entropies, brighter intrinsic luminosities, and stronger magnetic fields for young gas giants. This framework provides improved constraints for interpreting the masses and accretion rates of directly imaged exoplanets.}


   \maketitle
   \nolinenumbers   

\section{Introduction}

Following the discovery of 51 Pegasi b \citep{mayor1995}, elucidating the origins of exoplanetary systems has become a central objective in modern astronomy. Nevertheless, the exact mechanisms governing planet formation remain incompletely understood, largely due to the scarcity of observational constraints during the earliest evolutionary stages. In recent years, direct imaging \citep{bowler2016} has emerged as a powerful technique for identifying young, forming bodies. These detections rely primarily on high-contrast infrared observations, which capture the high intrinsic thermal luminosity of young, massive planets \citep{marois2008,marois2010,lagrange2010,macintosh2015,chauvin2017,hammond2023,vancapelleveen2025,bernardi2026}. Furthermore, actively accreting protoplanets can now be probed via \ce{H\alpha} emission \citep{keppler2018,haffert2019,zhou2021,close2025a,close2025b,zhou2025}, offering a diagnostic of magnetospheric accretion—a process that generally requires strong planetary magnetic fields (\citealp{batygin2025}; though see \citealp{chen2026b}).

Two primary paradigms account for the formation of gas giants. The first is the core-accretion model \citep{mizuno1980,pollack1996,Marley2007}, in which massive solid cores assemble via the coagulation of dust and planetesimals before accreting gas from the surrounding protoplanetary disk (PPD). More recently, this classical picture has been extended by pebble accretion theory \citep{ormel2010,michiel2012,johanson2017}, which alleviates the slow core-growth timescales and migration bottlenecks inherent to planetesimal accretion \citep{joanna2023}. The second paradigm is the gravitational instability model \citep{cameron1978,boss1997,nayakshin2010}, wherein gas giants form directly via the gravitational collapse of localized, self-gravitating clumps within the PPD. Gravitational instability naturally accounts for gas giant formation in low-metallicity environments \citep{simon2018}, operates effectively at wide orbital separations \citep{niyang2025}, and preferentially produces high-mass companions \citep{helled2014}.

Within the core-accretion framework, the favored formation pathway for Jupiter and Saturn, planets acquire the bulk of their mass during runaway gas accretion. In this regime, low-angular-momentum gas falls onto the protoplanet at near free-fall velocities, driving a strong accretion shock that radiates away a substantial fraction of the kinetic energy \citep{Marley2007,Berardo2017}. Because most of the planetary envelope passes through this shock front, the fraction of entropy retained in the post-shock gas fundamentally sets the initial thermal state of the young giant. Depending on the efficiency of this radiative cooling, the resulting hot- or warm-start states can amplify convection- and rotation-driven magnetic dipole fields \citep{eliaslopez2025}, regulate magnetospheric accretion from the circumplanetary disk (CPD) \citep{batygin2018}, shape satellite system architectures \citep{batygin2025,Fujii2026}, and promote the long-term survival of dilute cores \citep{vazan2018,sur2025,knierim2026}.

To characterize this shock physics, \citet{marleau2017,Marleau2019} employed one-dimensional radiation-hydrodynamic (RHD) simulations to map post-shock thermodynamic states across planetary masses and radii. \citet{Chen2022} extended this approach by incorporating a realistic equation of state (EoS), showing that \ce{H2} dissociation strongly modifies the post-shock entropy. Because \ce{H2} dissociation and temperature-dependent opacities introduce strong nonlinearities, mapping these thermodynamic regimes requires extensive parameter surveys. However, while one-dimensional RHD models capture the local shock microphysics, their computational expense precludes direct coupling to long-term evolutionary calculations.

In reality, runaway gas accretion is tightly coupled to the secular evolution of the planetary interior. Resolving both processes simultaneously from first principles is computationally prohibitive owing to the extreme separation of timescales between the planet's Kelvin--Helmholtz cooling timescale ($\tau_{\mathrm{KH}} \sim 10^{5}$\,yr) and the local free-fall accretion timescale ($\tau_{\mathrm{acc}} \sim 10^{-4}$\,yr). To bridge this gap, we train a physics-informed neural network (PINN) on a suite of RHD shock simulations to serve as a fast surrogate model for post-shock thermodynamic states, establishing a self-consistent pipeline that links short-timescale shock microphysics to long-term interior evolution. From the resulting interior structures, we can self-consistently derive the planet's intrinsic luminosity and an upper limit on its dynamo-generated magnetic field strength across arbitrary accretion histories.

The remainder of this paper is organized as follows. In Sect.~\ref{sec:methods}, we detail the physical formulation of the accretion shock model and the PINN architecture. We present the RHD simulation results in Sect.~\ref{sec:results_sim} and evaluate the performance of the PINN surrogate model in Sect.~\ref{sec:results_pinn}. In Sect.~\ref{sec:evolution}, we couple the RHD code \texttt{Guangqi} \citep{chen2026} with the Modules for Experiments in Stellar Astrophysics (\texttt{MESA}) via the trained surrogate model to construct self-consistent evolutionary tracks for young gas giants. Finally, we discuss the broader implications of our findings in Sect.~\ref{sec:discussion} and summarize our conclusions in Sect.~\ref{Sec:conclusion}.

\section{Methods} \label{sec:methods}

We first generate an extensive dataset of first-principles RHD simulations using {\tt Guangqi}, and subsequently use these data to train a fast physics-informed neural network (PINN) surrogate model for the post-shock thermodynamic state.

\subsection{First-principles calculations}
\subsubsection{Physical model}
\begin{figure}[h!]
    \centering
    \includegraphics[width=0.8\hsize]{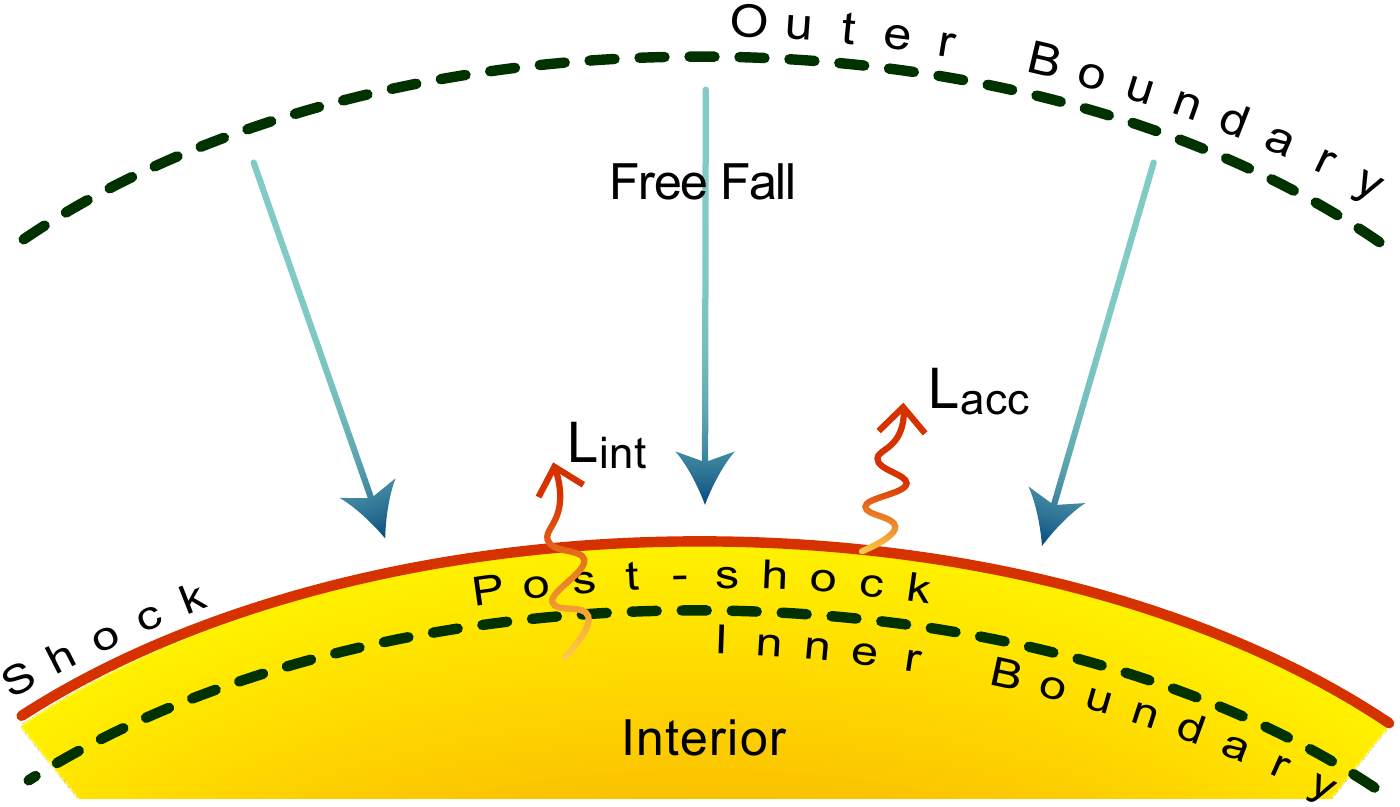}
    \caption{Schematic diagram of the hydrodynamic processes in our model. Gas infalls at free-fall velocity from the outer boundary of the computational domain (defined in Sect.~\ref{sec:grid}) and forms an accretion shock at the top of the planetary atmosphere, after which the shocked gas cools and settles onto deeper layers. Here, $L_{\mathrm{int}}$ and $L_{\mathrm{acc}}$ denote the internal luminosity from the planetary interior and the accretion luminosity, respectively.}
    \label{fig:schematic}
\end{figure}

Because the accretion flow evolves on a much longer timescale than the dynamical and shock-cooling timescales, we seek steady-state solutions. To this end, we adopt the radiation-hydrodynamic framework of \citet{Chen2022} and extend it to include the planet's internal luminosity, as illustrated schematically in Fig.~\ref{fig:schematic}. Formulated in a spherically symmetric laboratory frame, the governing equations combine hydrodynamics with flux-limited diffusion (FLD) radiation transport and a realistic EoS:
\begin{eqnarray}
	\pdv{\rho}{t}+\frac{1}{r^2}\pdv{(r^{2}\rho v)}{r}&=&0\ ,		\\
	\pdv{(\rho v)}{t}+\frac{1}{r^2}\pdv{(r^{2}\rho v^{2})}{r}&=&-\pdv{P}{r}-\rho\pdv{\phi}{r}\ ,	\\
	\pdv{E}{t}+\frac{1}{r^2}\pdv{[r^{2}(E+P)v]}{r}&=&\dot{Q}_{\mathrm{int}}+\mathbb{G}-\rho v\pdv{\phi}{r}\ ,	\label{eqn:energyhydro}\\
	\pdv{E_{\mathrm{r}}}{t}+\frac{1}{r^2}\pdv{(r^{2}F_{\mathrm{r}})}{r}&=&-\mathbb{G}\ ,	\label{eqn:energyrad} \\
    \phi&=&-G\mpl/r,  \\
    \mathbb{G}&=&\rho c\kappa_{\mathrm{P}}(E_{\mathrm{r}}-a_{\mathrm{rad}}T^{4}),   \\
    E&=&\rho[v^2/2+\epsilon_{\mathrm{g}}(\rho,T)],
\end{eqnarray}
where $\rho$, $v$, $P$, $T$, $E$, and $\epsilon_{\mathrm{g}}$ denote the gas density, radial velocity, pressure, temperature, total energy density, and specific internal energy, respectively, and $\phi,\mpl,G$, $c$, and $a_{\mathrm{rad}}$ are the gravitational potential, planet mass, gravitational constant, speed of light, and radiation constant. The gas composition is fixed at 74\% hydrogen and 26\% helium by mass. Whereas \citet{Chen2022} employed an EoS accounting for both \ce{H2} dissociation and hydrogen ionization, they demonstrated that ionization occurs only in very strong shocks. To accelerate our parameter survey, we neglect hydrogen ionization here; further details of this EoS are provided in Appendix~\ref{AA}.

We adopt the same opacity tables \citep{Semenov2003,Malygin2014}, FLD radiation transport scheme \citep{levermore1981}, and implicit solver \citep{petsc-user-ref,petsc-efficient} as in \citet{Chen2022}. Consequently, the radiation energy density ($E_{\mathrm{r}}$), radiation flux ($F_{\mathrm{r}}$), radiation--gas energy exchange rate ($\mathbb{G}$), Planck mean opacity ($\kappa_{\mathrm{P}}$), and Rosseland mean opacity ($\kappa_{\mathrm{R}}$) are evaluated identically to \citet{Chen2022}.

A key physical addition in this work is the planetary internal luminosity, modeled via the volumetric heating source term $\dot{Q}_{\mathrm{int}}$ in Eq.~(\ref{eqn:energyhydro}):
\begin{equation}
    \dot{Q}_{\mathrm{int}}=L_{\mathrm{int}}/dV_{3},
\end{equation}
where $dV_{3}$ is the combined volume of the three innermost spherical shells. Distributing $L_{\mathrm{int}}$ uniformly across this thin inner layer approximates a fixed basal radiative flux while avoiding numerical instabilities.\footnote{Imposing a high radiative flux directly at the inner boundary degrades the accuracy of the implicit solver and induces numerical oscillations.} Because this heating zone is geometrically thin relative to the atmospheric scale height and resides deep within the optically thick envelope, any error introduced by this treatment is negligible.

\subsubsection{Boundary and initial conditions}

The boundary conditions are configured to represent runaway gas accretion in the absence of external irradiation or prior assumptions regarding the deep planetary interior. Following \citet{Chen2022}, we impose the radiation boundary conditions
\begin{equation}
\begin{cases}
\pdv{E_{\mathrm{r}}}{r}=0 & \text{at } r=R_{\mathrm{in}}  \quad\text{(zero radiative flux)}, \\
\pdv{(E_{\mathrm{r}}r^{2})}{r}=0 & \text{at } r=R_{\mathrm{out}}  \quad\text{(optically thin)},
\end{cases}
\end{equation}
where $R_{\mathrm{in}}$ and $R_{\mathrm{out}}$ denote the inner and outer boundary radii, respectively.

At the hydrodynamic outer boundary, gas enters steadily at the local free-fall velocity for a prescribed mass accretion rate $\dot{M}$:
\begin{eqnarray}
	v_{\mathrm{out}}&=&-\sqrt{2G\mpl/R_{\mathrm{out}}}\ ,	\\
	\rho_{\mathrm{out}}&=&-\dot{M}/(4\pi R_{\mathrm{out}}^{2}v_{\mathrm{out}})\ ,\label{eq:rhoinj}\\
    \pdv{T_{\mathrm{out}}}{r}&=&0.
\end{eqnarray}
At the inner boundary, a reflective condition is maintained until the basal gas pressure reaches a specified target value. Thereafter, the primitive variables $(\rho, T, P)$ are held fixed and the ghost-cell velocity is set to zero, allowing gas to exit the domain whenever the mass flux predicted by the approximate HLLC Riemann solver \citep{chen2019} is negative. This treatment mimics a quasi-static envelope whose structure remains stationary over a free-fall timescale. Over secular evolutionary timescales, however, the thermodynamic state at $R_{\mathrm{in}}$ evolves, necessitating the coupled shock--interior framework developed in Sect.~\ref{sec:evolution}.

Because we consider only steady-state solutions, the final profiles are insensitive to the initial state provided the initial gas mass in the domain is negligible. We therefore initialize all runs with $\rho_{0}=10^{-13}\,\mathrm{g\,cm^{-3}}$, $T_{0}=200\,$K, and $v_{0}=0$.

\subsubsection{Computational grid} \label{sec:grid}

To resolve the sharp shock structure, we employ a geometrically stretched radial grid extending from $R_{\mathrm{in}}$ to $R_{\mathrm{out}} = 10\,R_{\mathrm{in}}$. The base grid comprises $N_{x} = 256$ cells with a constant geometric stretching ratio $\Delta x_{n+1}/\Delta x_n = 1.009$. We apply static mesh refinement (SMR) across the shock region, with each refinement level doubling the spatial resolution. We adjust the refinement level (five to six levels) according to $R_{\mathrm{in}}$ and $\mpl$, yielding a finest cell width between $12.0\,$km and $54.2\,$km. Because the atmospheric pressure scale height at the inner boundary,
\begin{equation}
    \frac{H}{\mathrm{km}}=591\left(\frac{T}{1800\,\mathrm{K}}\right)\left(\frac{R_{\mathrm{in}}}{1.5\,\rj}\right)^{2}\left(\frac{2.3}{\mu}\right)\left(\frac{1\,\mj}{\mpl}\right),
\end{equation}
scales as $R_{\mathrm{in}}^{2}/\mpl$, adjusting the resolution accordingly ensures that $H$ is consistently resolved by more than 16 cells near the inner boundary across our parameter survey.

\subsubsection{A representative simulation from {\tt Guangqi}}

\begin{figure*}[!h]
    \centering
    \includegraphics[width=\linewidth]{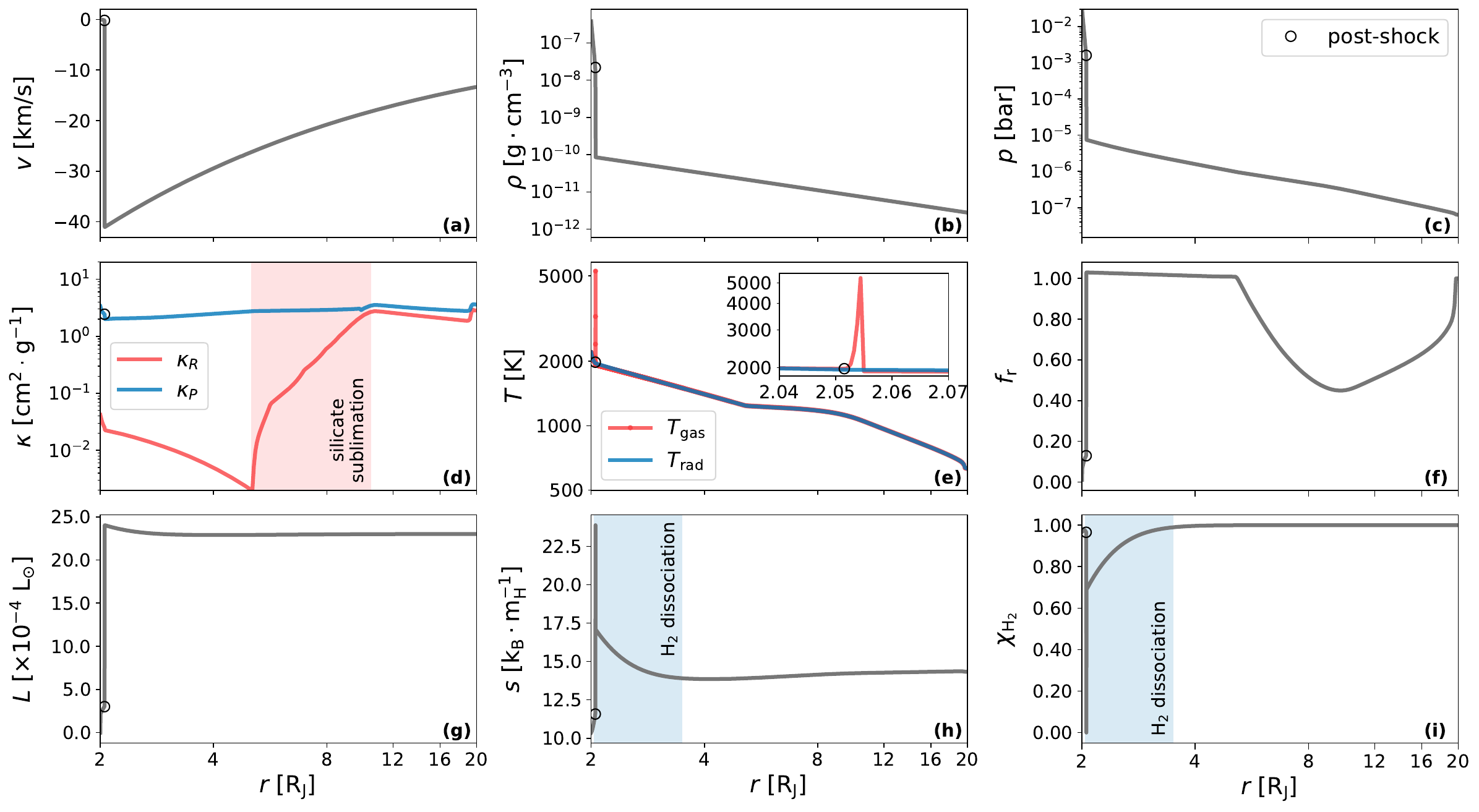}
    \caption{Representative steady-state accretion shock solution from {\tt Guangqi} for a gas giant with $\mpl = 1\,\mj$, $R_{\mathrm{in}}=2\,\rj$, $\dot{M} = 0.005\,\mathrm{\mathrm{M_\oplus}\,yr^{-1}}$, and $L_{\mathrm{int}}/L_{\mathrm{acc}} = 0.1$. Panels show radial profiles of (a) gas velocity, (b) density, (c) pressure, (d) Planck mean opacity (blue) and Rosseland mean opacity (red), (e) radiation temperature (blue) and gas temperature (red), (f) reduced radiation flux, (g) luminosity, (h) specific entropy, and (i) \ce{H2} mass fraction. Open circles mark the post-shock state.}
    \label{fig:guangqi_profile}%
\end{figure*}

Figure~\ref{fig:guangqi_profile} displays a representative steady-state solution for a $1\,\mj$ accreting protoplanet within a domain spanning $[2, 20]\,\rj$ at $\dot{M} = 0.005\,\mathrm{\mathrm{M_\oplus}\,yr^{-1}}$ and $L_{\mathrm{int}}/L_{\mathrm{acc}} = 0.1$. The infalling gas accelerates in free fall from the outer boundary (Fig.~\ref{fig:guangqi_profile}a) before decelerating abruptly at an accretion shock near $R_{\mathrm{in}}$. The flow requires $\sim 3 \times 10^5\,$s to build up an upper atmosphere and reach the target inner boundary pressure. We define the post-shock radius $\rpl$ as the location downstream of the Zel'dovich spike where the gas and radiation temperatures equilibrate (Fig.~\ref{fig:guangqi_profile}e), marked by open circles in each panel. As shown in Fig.~\ref{fig:guangqi_profile}i, radiative preheating from the shock drives substantial \ce{H2} dissociation in the upstream flow \citep{Chen2022}, leaving several percent of the hydrogen in atomic form in the post-shock region. Simultaneously, the intense accretion luminosity sublimates silicate grains, reducing the Rosseland mean opacity to $\kappa_{\mathrm{R}} \sim 10^{-2}\,\mathrm{cm^2\,g^{-1}}$ (Fig.~\ref{fig:guangqi_profile}d).

\subsection{Parameter survey with {\tt Guangqi}} \label{sec:results_sim}

We draw 1000 parameter combinations spanning the bulk of the physical regime relevant to runaway gas accretion: planetary mass $\mpl\in[0.1,3]\,\mj$, radius $R_{\rm in}\in[1,3.5]\,\rj$ \citep{spiegel2012,Hatzes2015,Bashi2017,Kubyshkina2022,thomas2025}, mass accretion rate $\dot{M}\in[10^{-4},10^{-2}]\,\mathrm{M_\oplus\,yr^{-1}}$ \citep{Mordasini2017,Lambrechts2019}, and internal luminosity $L_{\mathrm{int}}\in[10^{-6},10^{-3}]\,\mathrm{L_\odot}$ \citep{Marley2007,Mordasini2013}. Because cases with very low internal luminosity produce nearly indistinguishable post-shock states, we restrict our sample to $L_{\mathrm{int}}/L_{\mathrm{acc}}\ge 0.01$, where $L_{\mathrm{acc}} = G\mpl\dot{M}/R_{\mathrm{in}}$ is the accretion luminosity assuming complete conversion of gravitational potential energy into radiation. Below, we examine how each physical parameter governs the post-shock specific entropy, $s_{\mathrm{ps}}$.

\subsubsection{Planet mass, radius, accretion rate, and hydrogen dissociation}\label{sec:correlation}

\begin{figure*}
    \centering
    \includegraphics[width=\linewidth]{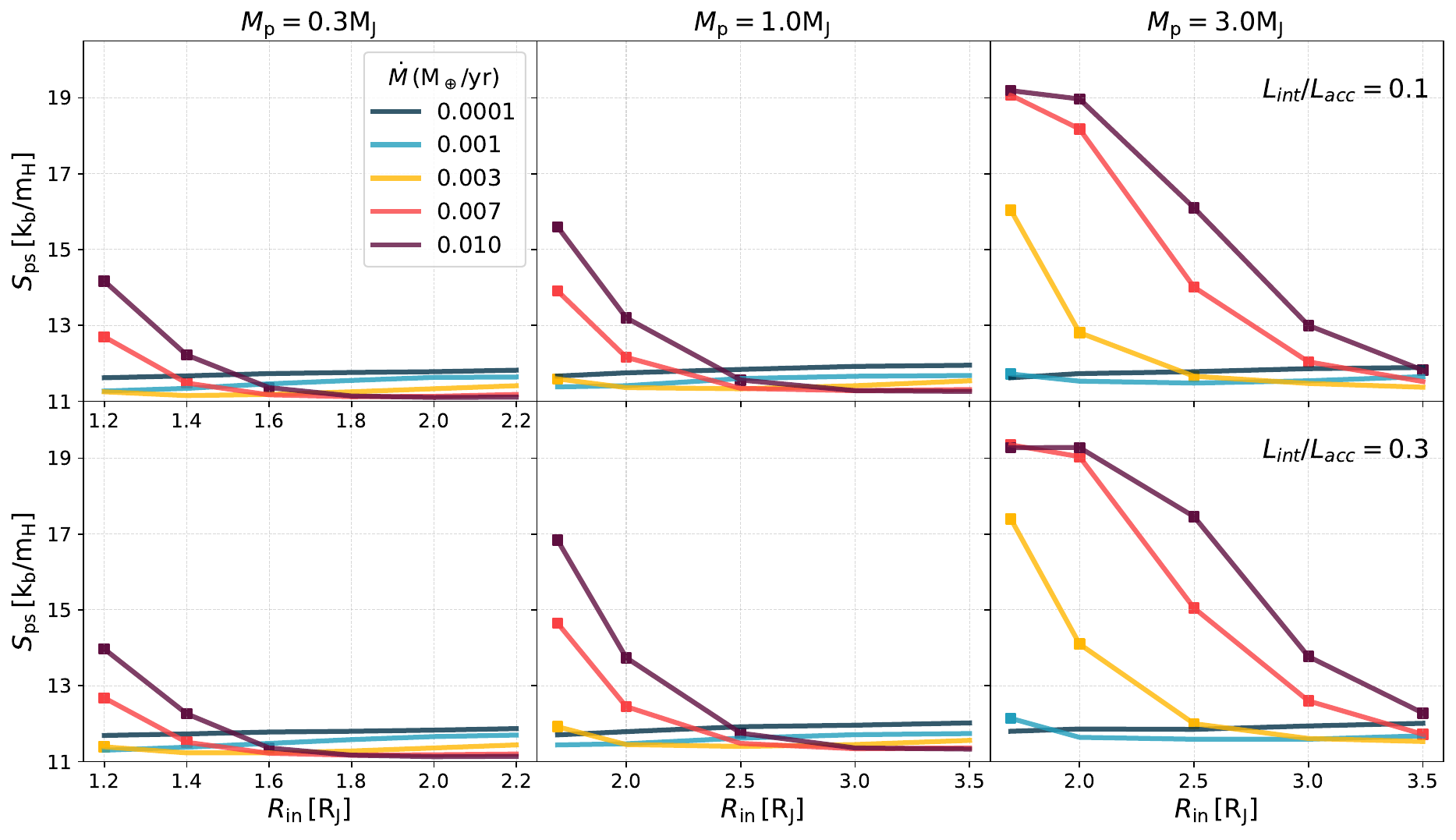}
    \caption{Post-shock specific entropy $s_{\mathrm{ps}}$ as a function of initial planetary radius $R_{\mathrm{in}}$ across varying accretion rates $\dot{M}$ (colors). Columns from left to right correspond to $\mpl=\{0.3,1,3\}\,\mj$, and rows correspond to $L_{\mathrm{int}}/L_{\mathrm{acc}}=\{0.1,0.3\}$. Filled squares denote simulations in which the post-shock atomic hydrogen mass fraction exceeds $\chi_{\mathrm{H,ps}}> 2\%$.}
    \label{fig:matrix}%
\end{figure*}

In the absence of \ce{H2} dissociation and internal molecular excitations, the specific entropy follows the ideal-gas expression:
\begin{align}
    s&=\frac{1}{\mu}\left(\frac{\gamma}{\gamma-1}\ln\frac{T}{T_0}-\ln\frac{P}{P_0}\right) \nonumber \\
    &=s_{0}+\frac{1}{\mu}\left(\frac{\gamma}{\gamma-1}\ln T-\ln P\right), \label{eqn:entropy}
\end{align}
where $\mu$ is the mean molecular weight, $\gamma$ is the adiabatic index, and $(s_0, T_0, P_0)$ defines a reference thermodynamic state. Because $P_{\mathrm{ps}}$ is set primarily by the inflow ram pressure while $T_{\mathrm{ps}}$ is regulated by radiative transport, both quantities follow approximate power-law scalings \citep[adapted from Eqs.~16 and 34b of][]{Marleau2019}:
\begin{eqnarray}
    P_{\mathrm{ps}}&=&\frac{\sqrt{2G}}{4\pi}\dot{M} \mpl^{1/2} \rpl^{-5/2},\label{eqn:analytic1}\\
    T_{\mathrm{ps}}&\approx&\left(\frac{2\sqrt{G}}{128\sqrt{2}\pi^{2}\sigma}\right)^{1/4}\kappa_{\mathrm{R}}^{1/4} \dot{M}^{1/2} \mpl^{1/8} \rpl^{-7/8},\label{eqn:analytic2}
\end{eqnarray}
where $\rpl$ and $\sigma$ are the post-shock radius and Stefan--Boltzmann constant. In this work, $R_{\rm in}<\rpl$ because $R_{\rm in}$ is at the base of the atmosphere while $\rpl$ is measured at the post-shock radius in {\tt Guangqi} simulations. Without \ce{H2} dissociation, $T_{\mathrm{ps}}$ typically remains below $1600\,$K. Within this regime, silicate dust sublimation drives a steep drop in $\kappa_{\mathrm{R}}$ with increasing temperature \citep{Semenov2003}.\footnote{Across the silicate sublimation window, $\kappa_{\mathrm{R}}$ drops by roughly three orders of magnitude over a $\sim 15\%$ increase in temperature, corresponding to a strongly negative effective power-law index.} Adopting a representative scaling $\kappa_{\mathrm{R}}\propto T^{-4}$ together with $\gamma=5/3$ (consistent with our EoS when \ce{H2} remains intact) yields
\begin{equation}
    s_{\mathrm{ps}}=s_{0}+\left(-\frac{3\ln\dot{M}}{8}-\frac{11\ln\mpl}{32}+\frac{45\ln\rpl}{32}\right)\bigg/\mu,\label{eqn:spscorrelation}
\end{equation}
where constant terms are absorbed into $s_0$. Equation~(\ref{eqn:spscorrelation}) predicts that $s_{\mathrm{ps}}$ correlates positively with $\rpl$ and negatively with both $\dot{M}$ and $\mpl$. Despite the strong nonlinearities of the full RHD system, the {\tt Guangqi} results in Fig.~\ref{fig:matrix} clearly recover two of these three trends.

In Fig.~\ref{fig:matrix}, simulations with significant \ce{H2} dissociation ($\chi_{\mathrm{H,ps}} > 2\%$) are marked by filled squares, whereas dissociation-free models are shown as circles. At fixed $\mpl$ and $\rpl$ within the dissociation-free regime (typically at large $\rpl$), $s_{\mathrm{ps}}$ decreases with increasing $\dot{M}$, in agreement with Eq.~(\ref{eqn:spscorrelation}). Similarly, at fixed $\dot{M}$ and $\mpl$, $s_{\mathrm{ps}}$ increases monotonically with $\rpl$. By contrast, no prominent negative correlation between $s_{\mathrm{ps}}$ and $\mpl$ is apparent across the parameter grid.

At high accretion luminosities, intense radiative heating raises the gas temperature to the \ce{H2} dissociation threshold, where each dissociating molecule absorbs $4.48\,$eV, producing a sharp jump in specific entropy. Consequently, the dissociated models (filled squares in Fig.~\ref{fig:matrix}) exhibit substantially higher $s_{\mathrm{ps}}$ than the undissociated models—a thermal dichotomy first identified by \citet{Chen2022}. Neglecting internal luminosity, the critical accretion rate $\dot{M}_{\mathrm{crit}}$ required to trigger \ce{H2} dissociation is approximately \citep{Chen2022}
\begin{equation}\label{eqn:mcrit}
    \frac{\dot{M}_{\mathrm{crit}}}{\mathrm{M_\oplus}\,\mathrm{yr}^{-1}}\approx7\times10^{-4}\bigg(\frac{\rpl}{\rj}\bigg)^{3}\bigg(\frac{\mpl}{\mj}\bigg)^{-1},
\end{equation}
assuming a geometric covering fraction $f_{\mathrm{c}}=1$ (adopted throughout this work), a reduced radiation flux $f_{\mathrm{r}}\approx 1$ following dust sublimation, and a global shock efficiency $\eta^{\mathrm{phys}}\approx 1$ (Eq.~\ref{eqn:eta}) prior to significant dissociation. Although a finite $L_{\mathrm{int}}$ lowers $\dot{M}_{\mathrm{crit}}$ by roughly $10\%$--$25\%$, we retain Eq.~(\ref{eqn:mcrit}) as a reference benchmark along the upper x-axis of Fig.~\ref{fig:ml_prediction} and as a reference contour in Fig.~\ref{fig:contour}.

\subsubsection{Internal luminosity}

Physically, $L_{\mathrm{int}}$ provides an additional basal heat flux that warms the post-shock region and increases $s_{\mathrm{ps}}$. Across our simulation grid, $s_{\mathrm{ps}}$ generally correlates positively with $L_{\mathrm{int}}$ when $\mpl$, $R_{\mathrm{in}}$, and $\dot{M}$ are held fixed.

In certain extreme regimes—typically involving partially dissociated post-shock gas—a very high $L_{\mathrm{int}}$ induces noticeable expansion of the outer envelope, which can cause $s_{\mathrm{ps}}$ to decrease slightly. An example appears in Fig.~\ref{fig:matrix} when comparing $L_{\mathrm{int}}/L_{\mathrm{acc}}=0.1$ and $0.3$ for $\mpl=0.3\,\mj$ and $\dot{M}=0.01\,\mathrm{M_\oplus}\,\mathrm{yr}^{-1}$. However, such conditions are unlikely to arise in nature, because low-mass planets cannot sustain high $L_{\mathrm{int}}/L_{\mathrm{acc}}$ ratios while accretion luminosity dominates \citep{Marleau2019}.

\subsubsection{Opacity}
\begin{figure}
    \centering
    \includegraphics[width=\hsize]{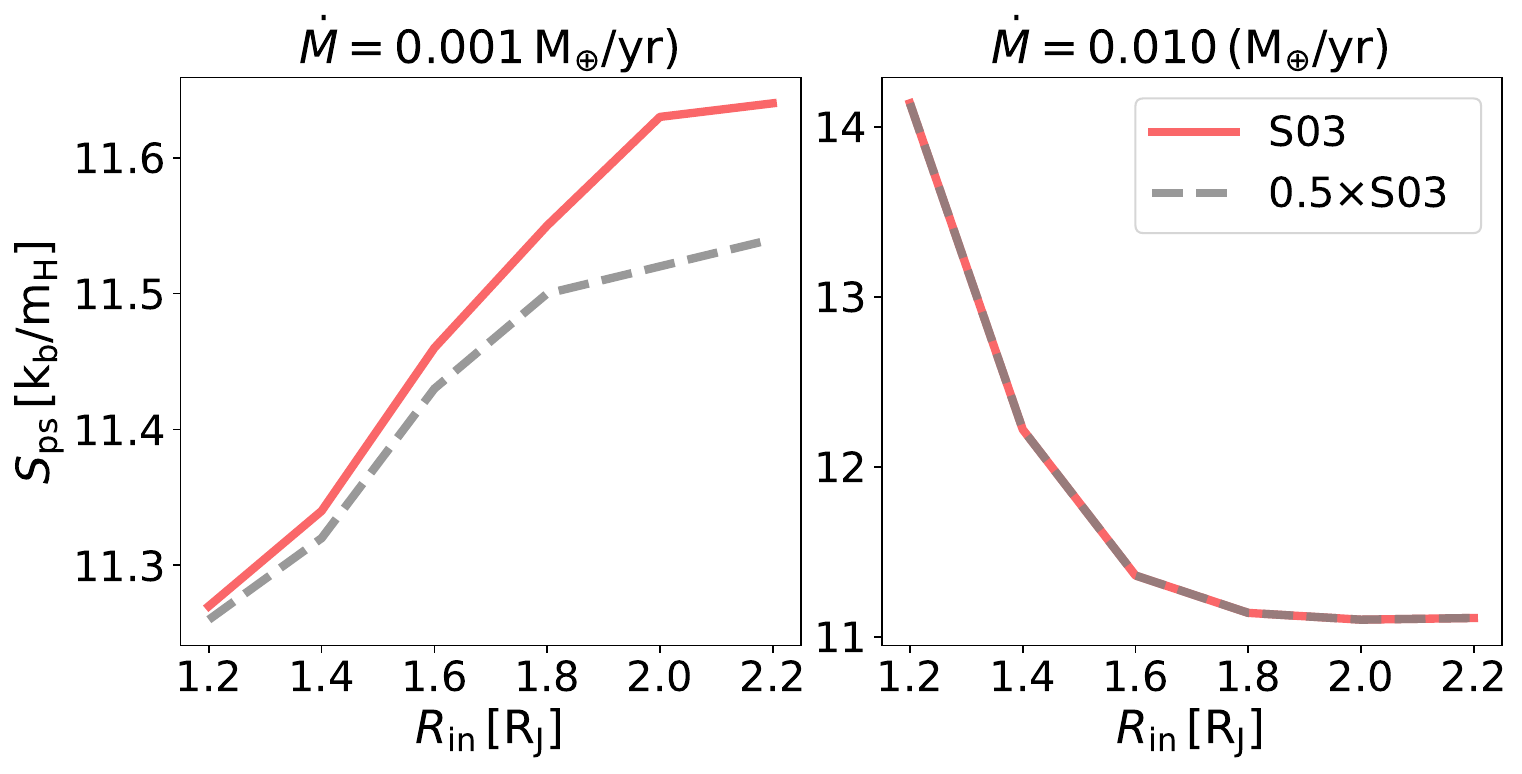}
    \caption{Effect of dust opacity on post-shock specific entropy for a $\mpl = 0.3\,\mj$ gas giant at $\dot{M} = 0.001\,\mathrm{M_\oplus\,yr^{-1}}$ (left) and $\dot{M} = 0.01\,\mathrm{M_\oplus\,yr^{-1}}$ (right). Solid lines adopt the fiducial dust opacity of \citet{Semenov2003}, whereas dashed lines adopt half that opacity with all other parameters fixed. The two curves overlap completely in the right panel.}
    \label{fig:opacity}%
\end{figure}

Dust opacity strongly influences the thermal evolution of young planets during their early formation stages \citep{piso2015}. To quantify its impact on $s_{\mathrm{ps}}$, we run two sets of simulations for a $\mpl = 0.3\,\mj$ planet at low ($\dot{M}=0.001\,\mathrm{M_\oplus\,yr^{-1}}$) and high ($\dot{M}=0.01\,\mathrm{M_\oplus\,yr^{-1}}$) accretion rates, comparing our fiducial opacity from \citet{Semenov2003} against a reduced opacity scaled by a factor of $0.5$ (Fig.~\ref{fig:opacity}).

At the low accretion rate (left panel), higher dust opacity yields a larger $s_{\mathrm{ps}}$, because the elevated optical depth impedes radiative cooling behind the shock. At the high accretion rate (right panel), however, shock heating raises temperatures above the dust sublimation threshold, rendering $s_{\mathrm{ps}}$ completely insensitive to the initial dust opacity.

\subsection{Machine learning algorithm}\label{sec:mlalgorithm}

\begin{figure}[h!]
    \centering
    \includegraphics[width=\hsize]{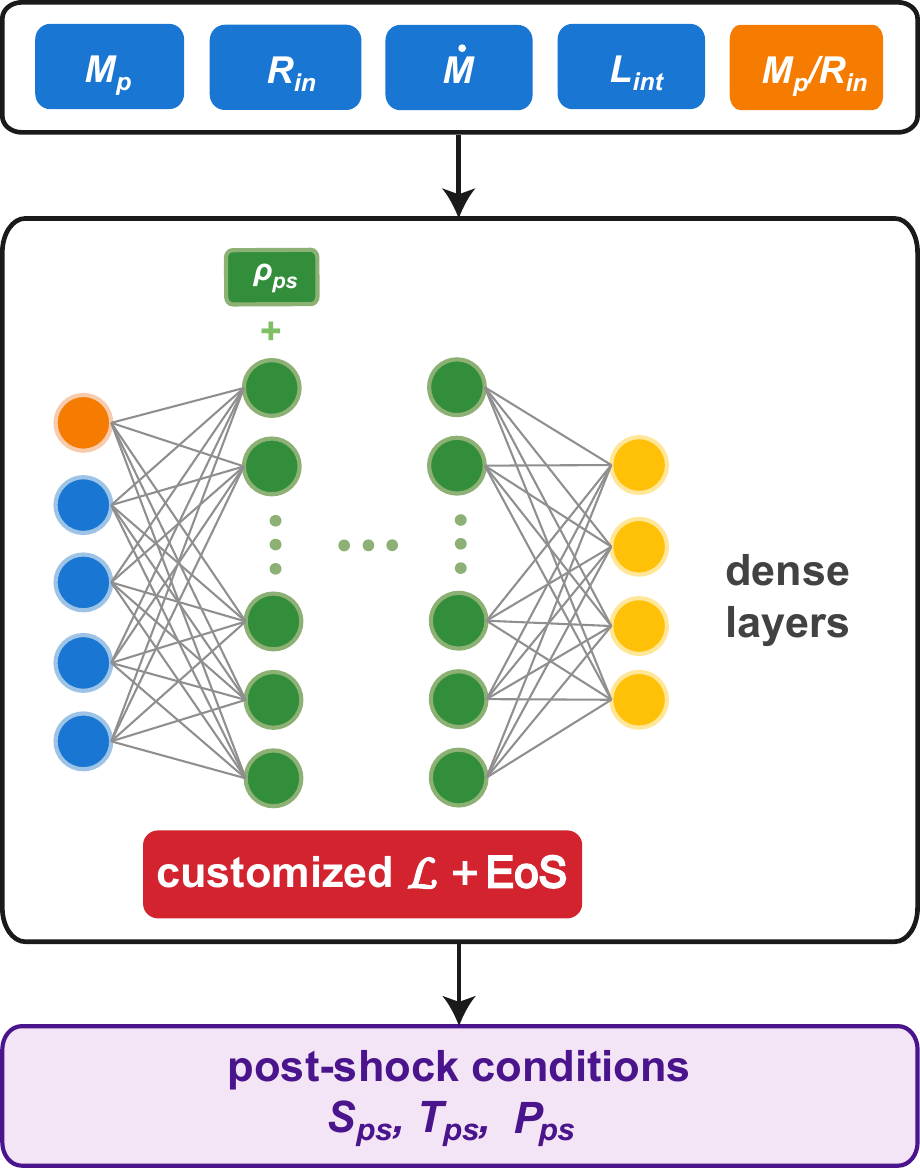}
    \caption{Architecture of the physics-informed neural network (PINN). Alongside the primary inputs, the physics-guided feature $\mpl/R_{\mathrm{in}}$ (orange) is supplied to the input layer. Within the dense hidden layers, the post-shock density $\rho_{\mathrm{ps}}$ (green) is incorporated as a supervised latent variable. A customized physics-informed loss function $\mathcal{L}$ (red) replaces the standard MSE objective. Depending on the EoS formulation in $\mathcal{L}$, the output layer returns $[s_{\mathrm{ps}}, T_{\mathrm{ps}}, P_{\mathrm{ps}}]$.}
    \label{fig:pinn_design}
\end{figure}

To enable rapid interpolation across our four-dimensional parameter space ($\mpl$, $R_{\mathrm{in}}$, $\dot{M}$, and $L_{\mathrm{int}}$), we construct a PINN surrogate model based on a multilayer perceptron (MLP) architecture (Fig.~\ref{fig:pinn_design}). The network outputs the post-shock specific entropy ($s_{\mathrm{ps}}$), temperature ($T_{\mathrm{ps}}$), and pressure ($P_{\mathrm{ps}}$), which are passed directly to \texttt{MESA} as outer boundary conditions for interior evolution calculations (Sect.~\ref{sec:evolution}).

Physical prior information and constraints are embedded into the model through the input features, the latent variables, and the loss function, respectively.
Each component is inspired by recent works from different disciplines that introduced physics-guided inputs \citep{huang2022,Willard2022}, latent variables incorporated into hidden layers \citep{pawar2021}, and physics-informed loss functions \citep{yu2020,ebert2021,nugroho2025}. 

First, we augment the four baseline inputs with the composite feature $\mpl/R_{\mathrm{in}}$, which scales directly with the specific gravitational potential energy released by the infalling gas and governs the overall accretion energy budget. Second, because the post-shock density $\rho_{\mathrm{ps}}$ is required to evaluate the thermodynamic relations in our loss function but is unknown prior to inference, we embed $\rho_{\mathrm{ps}}$ as a latent variable within a hidden layer. During training, this node is supervised against the simulated $\rho_{\mathrm{ps}}$ values, enabling the network to leverage density-dependent thermodynamic couplings during optimization without requiring post-shock quantities at inference time.

Third, rather than relying solely on a standard mean-squared-error (MSE) objective, we train the network using a composite loss function constrained by the EoS:
\begin{equation}
    \mathcal{L} = \mathbf{w} \cdot \mathbf{L} \, ,
    \label{lossFunction}
\end{equation}
where $\mathbf{w}$ is a vector of scalar weights and
\begin{equation}
\mathbf{L}=\left[L_{s},L_{T},L_{P},L_{\rho},L_{\mathrm{phy}},L_{\mathrm{con}}\right] \, .
\end{equation}
Here, $L_{s}$, $L_{T}$, $L_{P}$, and $L_{\rho}$ denote the standard MSE losses,
\begin{equation}
    \label{lmse}
    L_{\mathrm{mse}}(x_0, x) = \frac{1}{N} \sum_{i=1}^{N} \left(x_0^{(i)} - x^{(i)}\right)^2 ,
\end{equation}
evaluated over $N$ samples between the simulated ($x_0$) and predicted ($x$) values of $s_{\mathrm{ps}}$, $T_{\mathrm{ps}}$, $P_{\mathrm{ps}}$, and $\rho_{\mathrm{ps}}$, respectively.

The two physics-informed penalty terms, $L_{\mathrm{phy}}$ and $L_{\mathrm{con}}$, are defined as
\begin{equation}
\begin{aligned}
    L_{\mathrm{phy}} &= L_{\mathrm{mse}}(s_{\mathrm{ps,sim}},s(T_{\mathrm{ps,sim}},\rho_{\mathrm{ps,sim}})) \\
    & +L_{\mathrm{mse}}(P_{\mathrm{ps,sim}},P(T_{\mathrm{ps,sim}},\rho_{\mathrm{ps,sim}})) \, ,
\end{aligned}
\label{phyloss}
\end{equation}
\begin{equation}
\begin{aligned}
    L_{\mathrm{con}} &= L_{\mathrm{mse}}(s(T_{\mathrm{ps,sim}},\rho_{\mathrm{ps,sim}}),s(T_{\mathrm{ps,pre}},\rho_{\mathrm{ps,pre}})) \\
    & +L_{\mathrm{mse}}(P(T_{\mathrm{ps,sim}},\rho_{\mathrm{ps,sim}}),P(T_{\mathrm{ps,pre}},\rho_{\mathrm{ps,pre}})) \, .
\end{aligned}
\label{conloss}
\end{equation}
In both expressions, $s(T,\rho)$ and $P(T,\rho)$ are evaluated analytically via the EoS, the Saha equation, and hydrogen mass conservation (see Appendix~\ref{AA}). Although $L_{\mathrm{phy}}$ and $L_{\mathrm{con}}$ share a similar functional structure, they serve distinct roles: $L_{\mathrm{phy}}$ enforces thermodynamic consistency with the underlying EoS and Saha equilibrium on the training set, whereas $L_{\mathrm{con}}$ enforces structural consistency between the thermodynamic states implied by the simulated and predicted $(T_{\mathrm{ps}}, \rho_{\mathrm{ps}})$ pairs. Embedding physical structure across the inputs, latent representation, and loss function improves physical interpretability and mitigates overfitting on our 1000-model training set.

\subsection{PINN surrogate model} \label{sec:results_pinn}

\begin{figure}
    \centering
    \includegraphics[width=\hsize]{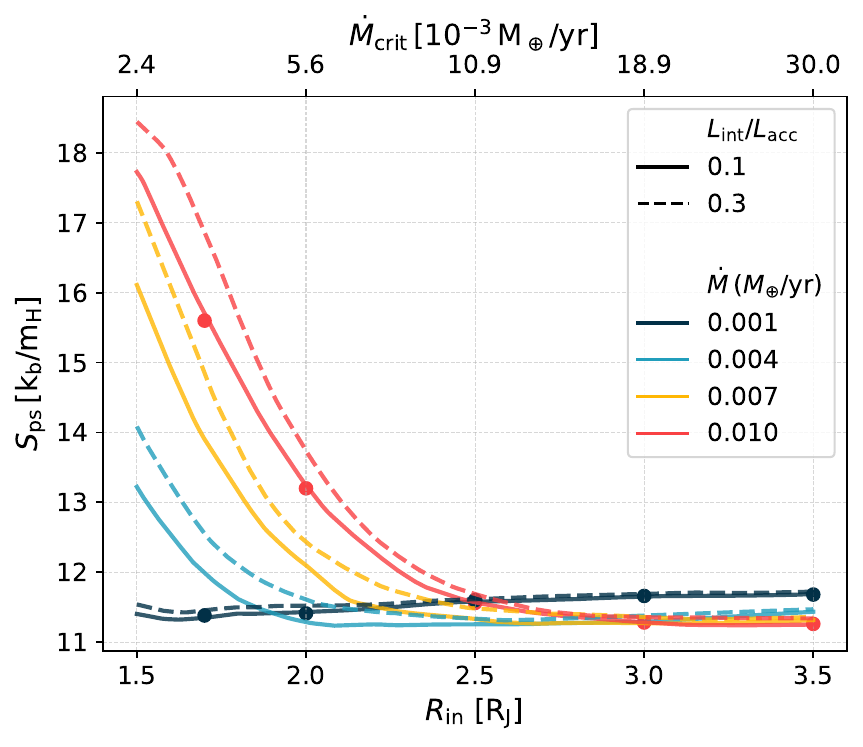}
    \caption{Comparison between post-shock specific entropy $s_{\mathrm{ps}}$ predicted by the PINN (curves) and measured directly from {\tt Guangqi} simulations (filled circles) for a $\mpl=1.0\,\mj$ planet. Curves span $\dot{M}=\{0.001,0.004,0.007,0.01\}\,\mathrm{M_\oplus}\,\mathrm{yr}^{-1}$ (colors) for $L_{\mathrm{int}}/L_{\mathrm{acc}} = 0.1$ (solid) and $0.3$ (dashed). Simulation points correspond to $L_{\mathrm{int}}/L_{\mathrm{acc}} = 0.1$ and $\dot{M}=\{0.001,0.01\}\,\mathrm{M_\oplus}\,\mathrm{yr}^{-1}$. The upper x-axis marks the critical accretion rate for \ce{H2} dissociation from Eq.~(\ref{eqn:mcrit}).}
    \label{fig:ml_prediction}
\end{figure}

\begin{figure*}
    \centering
    \includegraphics[width=\linewidth]{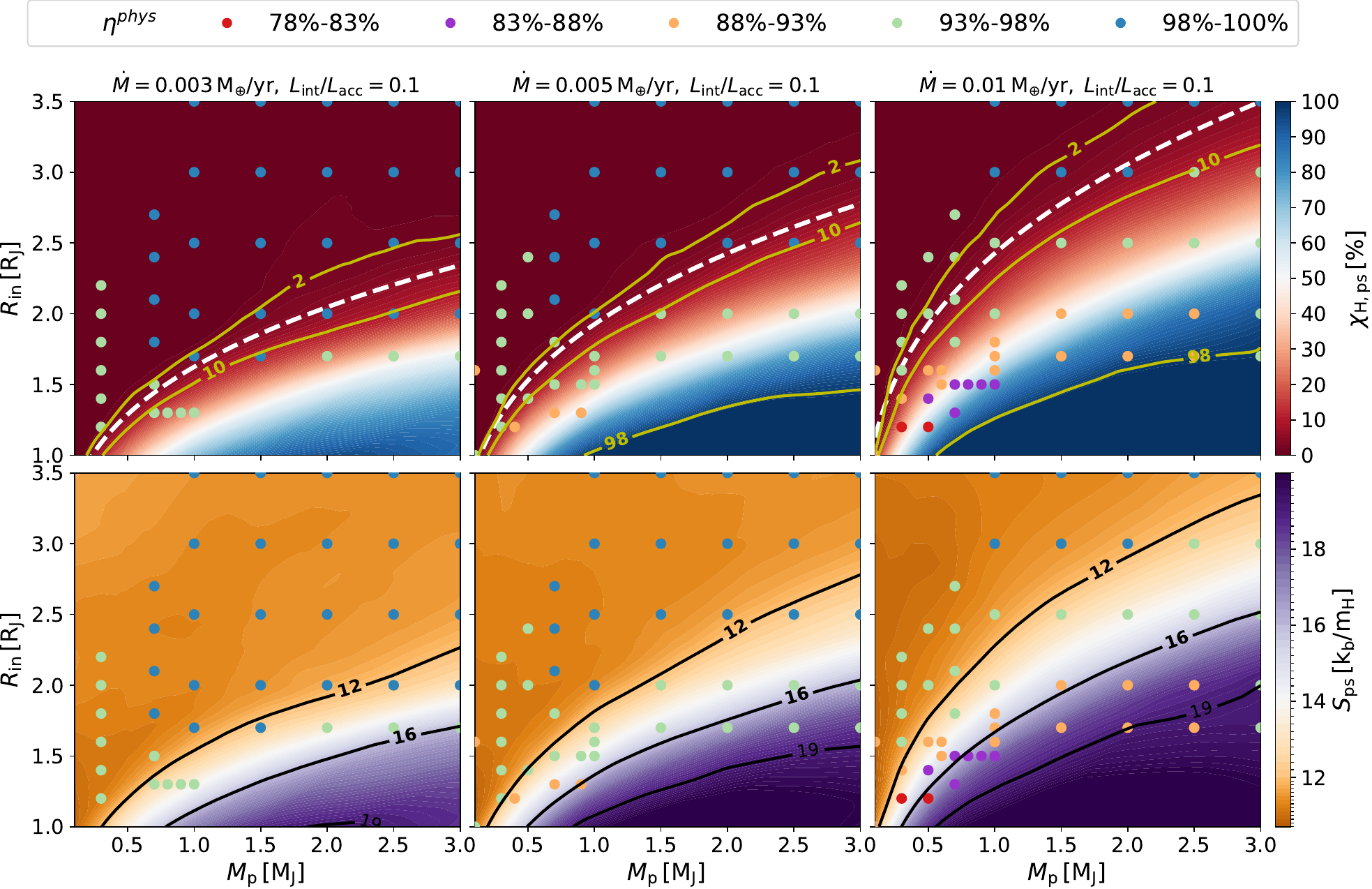}
    \caption{PINN-predicted post-shock atomic hydrogen mass fraction $\chi_{\mathrm{H,ps}}$ (top row) and specific entropy $s_{\mathrm{ps}}$ (bottom row) across the $(\mpl, R_{\mathrm{in}})$ parameter space for $\dot{M}=\{0.003,0.005,0.01\}\,\mathrm{M_\oplus}\,\mathrm{yr}^{-1}$ (left to right) at fixed $L_{\mathrm{int}}/L_{\mathrm{acc}}=0.1$. Top panels show contours of $\chi_{\mathrm{H,ps}} = \{2, 10, 98\}\%$ (black) alongside the analytical dissociation threshold from Eq.~(\ref{eqn:mcrit}) (white dashed). Bottom panels show contours of $s_{\mathrm{ps}} = \{12, 16, 19\}\,k_{\mathrm{B}}\,m_{\mathrm{H}}^{-1}$. Overlaid circles indicate the global shock efficiency $\eta^{\mathrm{phys}}$ (Eq.~\ref{eqn:eta}) measured in {\tt Guangqi} simulations: red ($78\%$--$83\%$), purple ($83\%$--$88\%$), orange ($88\%$--$93\%$), green ($93\%$--$98\%$), and blue ($98\%$--$100\%$).}
    \label{fig:contour}
\end{figure*}

Trained on the 1000 {\tt Guangqi} simulations, the PINN surrogate model evaluates post-shock thermodynamic states in $<10^{-4}\,$s per query while achieving a coefficient of determination $\mathcal{R}^{2}>0.99$ (Appendix~\ref{A:pinn_training}). Figure~\ref{fig:ml_prediction} illustrates its accuracy for a $1\,\mj$ planet against direct {\tt Guangqi} benchmarks. The surrogate model closely reproduces both thermodynamic regimes: in the dissociation-free regime, $s_{\mathrm{ps}}$ decreases with $\dot{M}$ and increases with $R_{\mathrm{in}}$ as expected from Sect.~\ref{sec:correlation}, whereas across the \ce{H2} dissociation boundary, the network accurately captures the sharp entropy jump.

Because the surrogate model generalizes continuously across the parameter range defined in Sect.~\ref{sec:results_sim}, we use it to construct high-resolution lookup tables sampled at $\Delta \mpl=10^{-2}\,\mj$ and $\Delta R_{\mathrm{in}}=10^{-2}\,\rj$. Figure~\ref{fig:contour} maps $\chi_{\mathrm{H,ps}}$ and $s_{\mathrm{ps}}$ across the $(\mpl, R_{\mathrm{in}})$ plane for three accretion rates.

In the top panels of Fig.~\ref{fig:contour}, the analytical dissociation threshold of Eq.~(\ref{eqn:mcrit}) lies between the $\chi_{\mathrm{H,ps}}=2\%$ and $10\%$ contours, providing a reasonable estimate for the onset of hydrogen dissociation. In the bottom panels, the accompanying entropy rise closely tracks dissociation, with the $s_{\mathrm{ps}}=12$ and $19\,k_{\mathrm{B}}\,m_{\mathrm{H}}^{-1}$ contours closely matching the $\chi_{\mathrm{H,ps}}=10\%$ and $90\%$ levels, respectively.

Finally, we overlay the global shock efficiency $\eta^{\mathrm{phys}}$ measured from {\tt Guangqi} as colored circles in Fig.~\ref{fig:contour}. Defined as the fraction of accretion energy radiated away by the shock \citep{marleau2017,Chen2022}, $\eta^{\mathrm{phys}}$ is given by
\begin{equation}\label{eqn:eta}
    \eta^{\mathrm{phys}} = \frac{\dot{E}(R_{\mathrm{out}})-\dot{E}(\rpl)}{\dot{E}(R_{\mathrm{out}})-\dot{E}(\rpl)+\dot{M}\left[\epsilon_{\mathrm{g,ps}}+v_{\mathrm{ps}}^{2}/2\right]} \, ,
\end{equation}
where $\dot{E}(r) = -\dot{M} \,(\epsilon_{\mathrm{g}}+v^2/2+P/\rho+\phi)$, evaluated between the outer boundary $R_{\mathrm{out}}$ and the post-shock radius $\rpl$, and the specific internal energy $\epsilon_{\mathrm{g,ps}}$ and gas velocity $v_{\mathrm{ps}}$ are evaluated at $\rpl$. Across all 1000 simulations in our survey, $\eta^{\mathrm{phys}}$ spans $78.5\%$ to $99.8\%$. Lower global shock efficiencies occur systematically when \ce{H2} dissociation absorbs the $4.48\,$eV binding energy per molecule, diverting a substantial fraction of the infalling gravitational potential energy into internal chemical energy rather than emergent radiation. 
For massive planets, the gravitational potential energy budget is given by
\begin{equation}
    \frac{G m_{\ce{H2}}\mpl}{\rpl}\approx55.5\left(\frac{\mpl}{3\,\mj}\right)\left(\frac{2\,\rj}{\rpl}\right)\,\mathrm{eV},
\end{equation}
which can be much larger than the \ce{H2} binding energy ($4.48\,$eV), driving $\eta^{\mathrm{phys}}$ close to 1. However, the \ce{H2} binding energy is still much larger than the mean thermal energy per particle ($\sim 0.2\,$eV) at the dissociation temperature; therefore, even partial dissociation can significantly increase the initial energy and entropy of newly formed gas giants.

\section{Application to the primordial states of gas giants with {\tt MESA}} \label{sec:evolution}

To assess the impact of \ce{H2} dissociation on gas giant formation, we model the accretion and long-term interior evolution of young planets using \texttt{MESA} r10398 \citep{mesa2011,mesa2013,mesa2015,mesa2018,paxton2019,mesa2023}, following the methodology of \citet{Berardo2016}.

\subsection{Initial and boundary conditions}\label{sec:initconditions}

\begin{table}
    \caption{Prescribed runaway gas accretion histories for the three representative planetary mass models. The accretion duration $t_{\mathrm{acc}}$ is kept short relative to the disk lifetime to represent the rapid runaway accretion phase.}
    \label{tab:accretion}
    \centering
    \begin{tabular}{lccc}
        \hline\hline
         & Saturn-like & Jupiter-like & Super-Jupiter \\
        \hline
        $M_{\rm final}$ [$\mj$] & 0.3 & 1.0 & 3.0 \\
        $t_{\mathrm{acc}}$ [Myr] & 0.141 & 0.237 & 0.451 \\
        $\dot{M}_{\mathrm{min}}$ [$\mathrm{M_\oplus\,yr^{-1}}$] & 0.0002 & 0.0003 & 0.0005 \\
        $\dot{M}_{\mathrm{max}}$ [$\mathrm{M_\oplus\,yr^{-1}}$] & 0.002 & 0.005 & 0.01 \\
        $M_{\mathrm{p,0}}$ [$\mj$] & 0.126 & 0.2 & 0.2 \\
        $M_{\mathrm{core}}$ [$\mj$] & 0.063 & 0.1 & 0.1 \\
        \hline
    \end{tabular}
\end{table}

We construct the initial planetary models using the \texttt{make\_planet} test suite in \texttt{MESA}. Each simulation starts from a fully convective gaseous sphere of initial mass $M_{\mathrm{p,0}}$ and radius $3.5\,\rj$. Following \citet{mesa2013}, we gradually replace the innermost mass $M_{\mathrm{core}}$ of the envelope with an inert core of fixed density $10\,\mathrm{g\,cm^{-3}}$ \citep{batygin2025}, and subsequently relax the planet for $\sim 10^3\,$yr to establish thermal and hydrostatic equilibrium.

Runaway gas accretion is implemented via the atmosphere module in \texttt{run\_star\_extras.f90}, which deposits newly accreted material onto the top of the envelope with a prescribed thermodynamic state. We compare two distinct outer boundary prescriptions:
\begin{enumerate}
    \item \emph{Photospheric} (cold-start baseline): Uses \texttt{MESA}'s default photospheric boundary condition, assigning the accreted gas the local photospheric specific entropy.
    \item \emph{Realistic EoS}: Replaces the photospheric boundary with the post-shock state $(P_{\mathrm{ps}}, T_{\mathrm{ps}})$ returned by our PINN surrogate model for the instantaneous values of $\mpl$, $\rpl$, $\dot{M}$, and $L_{\mathrm{int}}$ (Sects.~\ref{sec:mlalgorithm} and \ref{sec:results_pinn}), depositing accreted gas at the corresponding post-shock entropy $s_{\mathrm{ps}}$.
\end{enumerate}

The mass accretion rate is parameterized as a Gaussian function of instantaneous planetary mass $\mpl$:
\begin{equation}
    \dot{M}(\mpl) = \dot{M}_{\mathrm{min}} + (\dot{M}_{\mathrm{max}} - \dot{M}_{\mathrm{min}})\,
    \exp\left[-\frac{(\mpl-(M_{\mathrm{p,0}}+M_{\rm final})/2)^2}{2 [ (M_{\rm final}-M_{\mathrm{p,0}})/6 ] ^2}\right] \, ,
    \label{eq:mdot}
\end{equation}
which is symmetric about both the mass midpoint and the temporal midpoint of the accretion phase. As a proof of concept, we model the runaway growth of three representative planets reaching final masses of $M_{\rm final}=\{0.3, 1.0, 3.0\}\,\mj$, using the accretion parameters listed in Table~\ref{tab:accretion}. Throughout all simulations, the accreted gas is assumed to have a solar composition of $X=0.74$, $Y=0.246$, and $Z=0.014$ \citep{asplund2021}. Although the {\tt Guangqi} shock simulations adopt a metal-free H/He composition ($X=0.74$, $Y=0.26$) in the EoS, applying the resulting PINN boundary conditions to solar-metallicity {\tt MESA} envelopes introduces small error, because hydrogen dominates the gas internal energy and dissociation energetics while metals contribute only a minor fraction.

\subsection{Internal evolution}\label{sec:internalevolution}

\begin{figure*}[!h]
    \centering
    \includegraphics[width=1\linewidth]{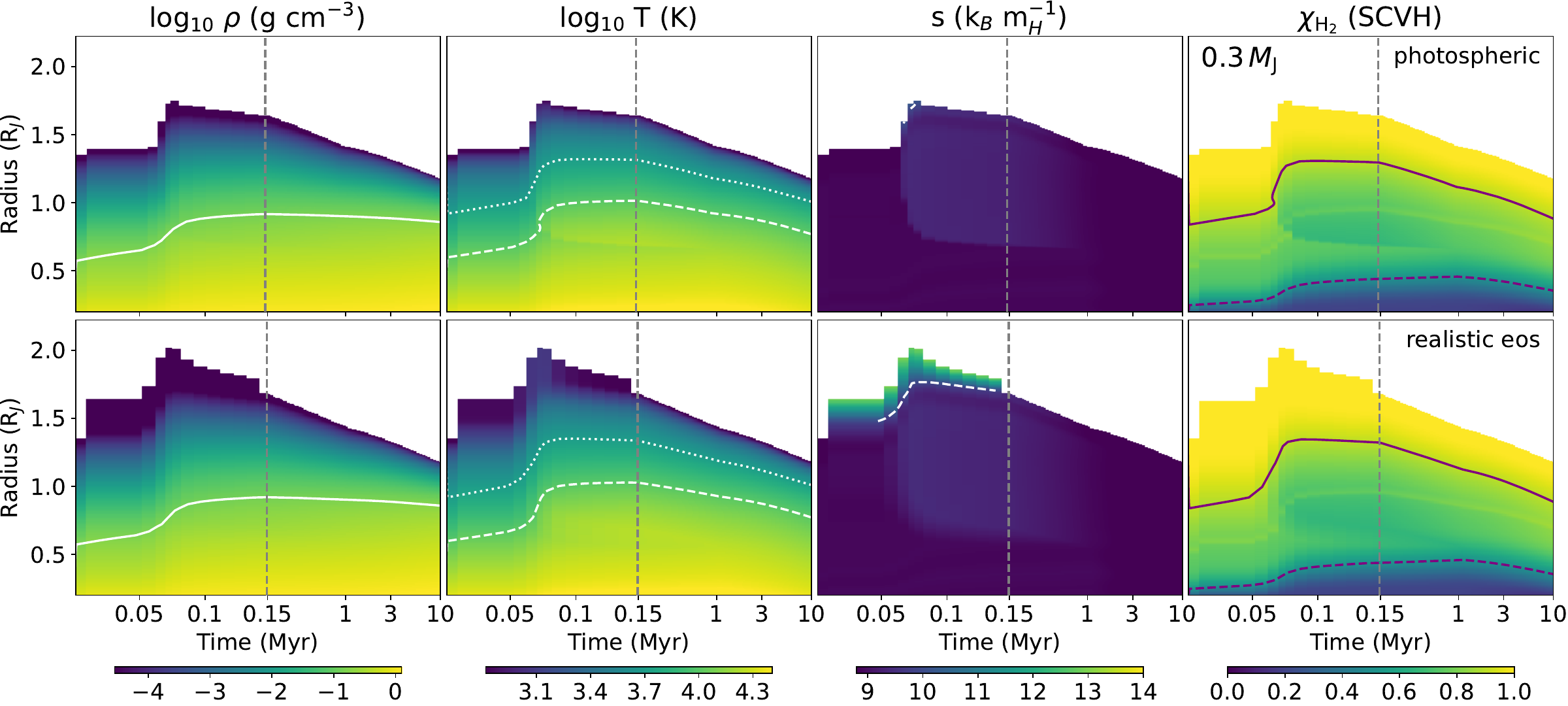}
    \caption{Interior structural evolution of the $M_{\rm final}=0.3\,\mj$ model. Columns from left to right show radial profiles of density, temperature, specific entropy, and molecular hydrogen mass fraction ($\chi_{\ce{H2}}$) over time. Top and bottom rows correspond to the ``photospheric'' and ``realistic EoS'' outer boundary conditions (Sect.~\ref{sec:initconditions}), respectively. Vertical grey dashed lines mark the end of runaway accretion ($t_{\mathrm{acc}}$). White solid contours in the density and entropy panels indicate $P=1\,$Mbar, denoting the approximate threshold for hydrogen metallization. White dotted and dashed contours in the temperature panels mark $T=\{5000, 10000\}\,$K, respectively. White dashed contours in the entropy panels trace $s=10\,k_{\mathrm{B}}\,m_{\mathrm{H}}^{-1}$. Purple solid and dashed contours in the $\chi_{\ce{H2}}$ panels denote $\chi_{\ce{H2}}=\{0.9, 0.5\}$, respectively.}
    \label{fig:0.3mj}
\end{figure*}

\begin{figure*}[!h]
    \centering
    \includegraphics[width=1\linewidth]{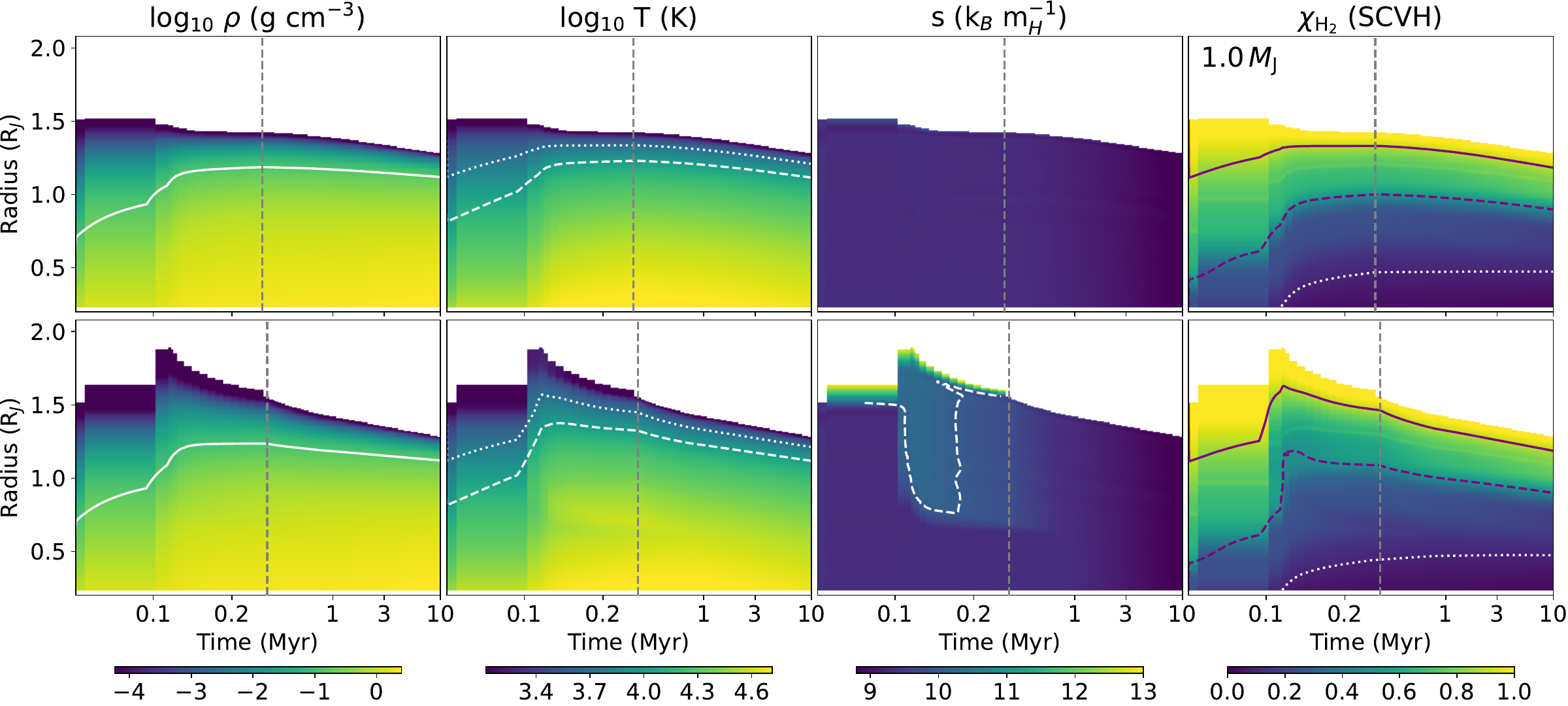}
    \caption{Same layout as Fig.~\ref{fig:0.3mj}, but for the $M_{\rm final}=1.0\,\mj$ model. White dotted contours in the $\chi_{\ce{H2}}$ panels indicate $\chi_{\ce{H2}}=0.1$.}
    \label{fig:1mj}
\end{figure*}

\begin{figure*}[!h]
    \centering
    \includegraphics[width=1\linewidth]{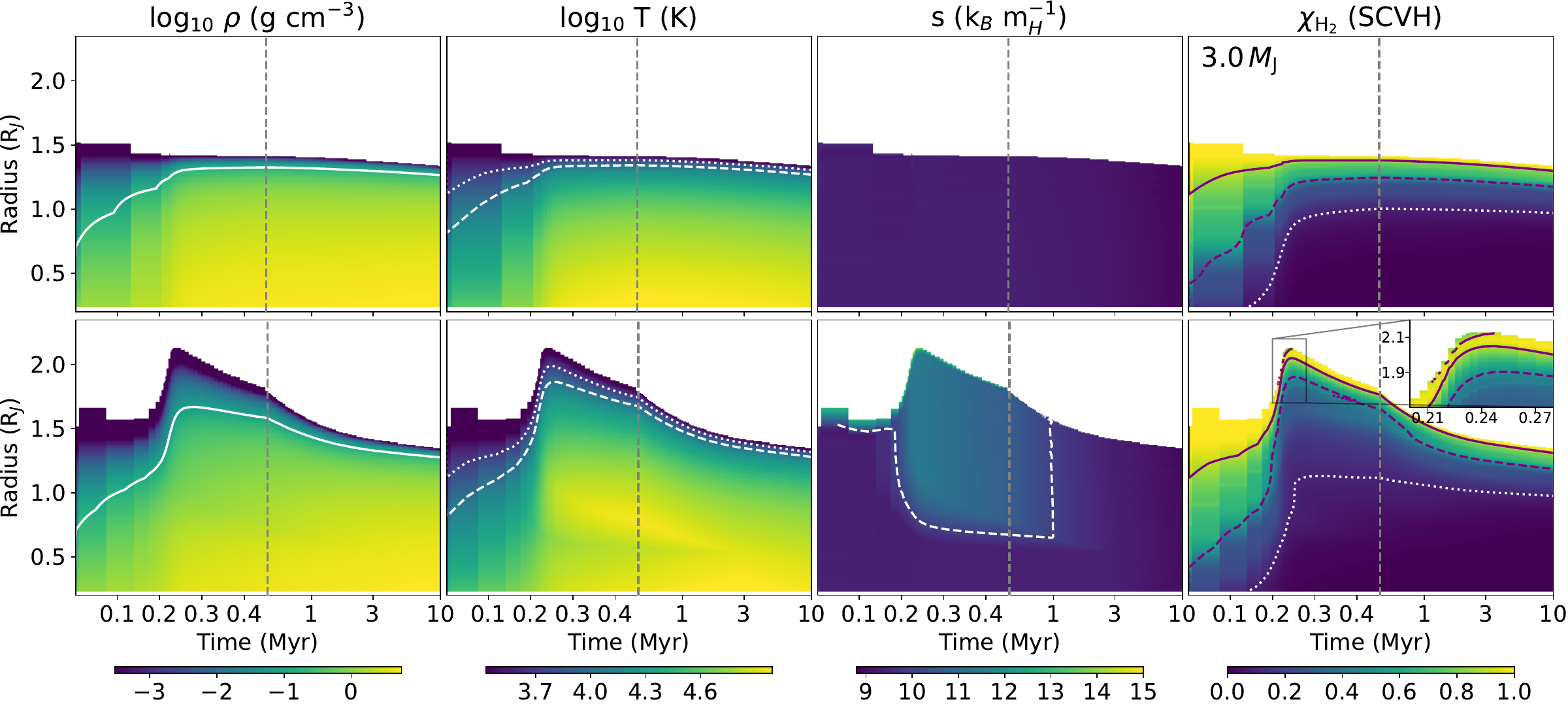}
    \caption{Same layout as Fig.~\ref{fig:1mj}, but for the $M_{\rm final}=3.0\,\mj$ model. In the zoomed-in panel, partial \ce{H2} dissociation occurs in the top layer of the envelope, set by the PINN surrogate model.}
    \label{fig:3mj}
\end{figure*}

Figures~\ref{fig:0.3mj}--\ref{fig:3mj} compare the interior structural evolution under the ``photospheric'' and ``realistic EoS'' outer boundary conditions for the $0.3\,\mj$, $1.0\,\mj$, and $3.0\,\mj$ models, respectively.

For the lowest-mass case ($M_{\rm final}=0.3\,\mj$; Fig.~\ref{fig:0.3mj}), the interior profiles remain nearly identical between the two boundary prescriptions. At $1.0\,\mj$ and $3.0\,\mj$, however, substantial structural differences emerge (Figs.~\ref{fig:1mj} and \ref{fig:3mj}). Under the ``realistic EoS'' boundary condition, the more massive planets develop markedly hotter, higher-entropy interiors, as reflected by the outward shift of the $s=10\,k_{\mathrm{B}}\,m_{\mathrm{H}}^{-1}$ contour. This divergence arises from two effects: first, the realistic post-shock boundary retains higher entropy than the photospheric approximation even prior to dissociation; second, once post-shock temperatures exceed the \ce{H2} dissociation threshold, the specific entropy rises sharply. Because only sufficiently massive planets generate the gravitational potential energy needed to dissociate \ce{H2} at the shock front, high-entropy envelopes naturally emerge at $M_{\rm final}\ge 1.0\,\mj$.

In the fourth column of Figs.~\ref{fig:0.3mj}--\ref{fig:3mj}, $\chi_{\ce{H2}}$ is evaluated using the Saumon--Chabrier--van Horn (SCVH) EoS \citep{saumon1995}. The decline of $\chi_{\ce{H2}}$ toward the planetary center reflects pressure-induced ionization and hydrogen metallization at high densities. Because the SCVH EoS is known to overestimate the hydrogen metallization pressure—expected to occur at $\sim 1\,$Mbar under the conditions of young gas giants \citep{hinz2020}—we adopt the $P=1\,$Mbar surface as the boundary of the metallic dynamo region in Sect.~\ref{sec:dynamo}. Employing the newer Chabrier--Mazevet--Soubiran (CMS) EoS \citep{chabrier2019} in updated versions of \texttt{MESA} would capture this transition more accurately.

Figure~\ref{fig:B_L}a shows the resulting intrinsic luminosity evolution for all three planetary masses, with filled circles marking the end of runaway accretion. Planets evolved with the PINN-predicted ``realistic EoS'' boundary condition are roughly $4$--$10$ times more luminous than their ``photospheric'' counterparts. Consistent with the interior structural trends, this luminosity enhancement is strongest for $M_{\rm final} \ge 1.0\,\mj$, where post-shock \ce{H2} dissociation injects substantially higher entropy into the accreting envelope \citep{Chen2022}.

\subsection{Dynamo-induced magnetic field strength}\label{sec:dynamo}

\begin{figure*}[!h]
    \centering
    \includegraphics[width=1\linewidth]{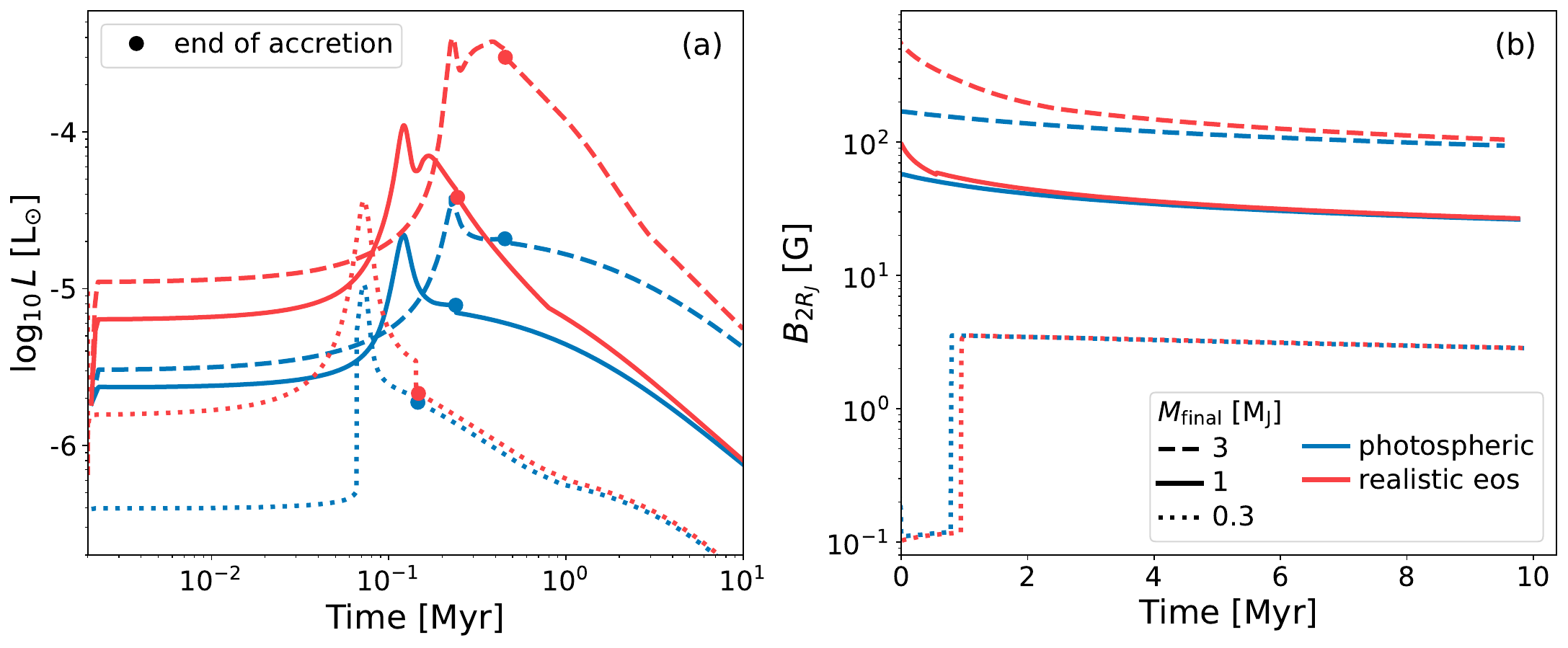}
    \caption{Evolution of (a) intrinsic luminosity and (b) dipole magnetic field strength at $2\,\rj$ in the equatorial plane over the first $10\,$Myr for the $0.3\,\mj$ (dotted), $1.0\,\mj$ (solid), and $3.0\,\mj$ (dashed) models. Blue curves represent the default ``photospheric'' \texttt{MESA} models, whereas red curves show the ``realistic EoS'' models incorporating post-shock boundary conditions from {\tt Guangqi}. Filled circles in panel (a) denote the end of runaway accretion, and panel (b) displays the magnetic field evolution following the completion of accretion.}
    \label{fig:B_L}
\end{figure*}

The magnetic field strength of a young gas giant depends primarily on its convective energy flux and rotation rate \citep{christensen2009}. Because our one-dimensional models do not evolve planetary spin, we assume rapid rotation in the regime where dynamo generation is independent of spin, thereby establishing an upper limit on the dipole field strength.

In planetary dynamo theory, efficient magnetic field generation requires high electrical conductivity \citep{brandenburg2005}. In gas giants, electrical conductivity is dominated by metallic and pressure-ionized hydrogen at pressures $P\gtrsim 1\,$Mbar \citep{french2012,hinz2020}. Following \citet{Fujii2026}, we therefore evaluate dynamo generation within the convective region at $P > 1\,$Mbar using the energy-flux scaling law of \citet{christensen2009}:
\begin{equation}
    \frac{\langle B^{2}\rangle}{2\mu_0}=\bar{c}\langle\rho\rangle^{1/3}\left(Fq_0\right)^{2/3},
\end{equation}
where $\mu_0$ is the vacuum magnetic permeability, $\langle B^2\rangle$ is the mean-square magnetic field strength in the dynamo region, $\langle\rho\rangle$ is the mean density of that region, $q_0$ is the convective energy flux at the outer boundary of the dynamo zone ($R_{\mathrm{dynamo}}$, defined by $P=1\,$Mbar), and $F$ is a geometric factor accounting for radial variations in flux and density. Following \citet{Fujii2026}, we set the dimensionless efficiency coefficient to $\bar{c} = 0.63$.

Assuming a dipolar geometry, the surface-extrapolated equatorial magnetic field strength at $r = 2\,\rj$ is given by
\begin{equation}
    B_{2\rj}=\frac{\sqrt{\langle B^{2}\rangle}}{2\sqrt{2}}\left(\frac{R_{\mathrm{dynamo}}}{2\rj}\right)^{3},
\end{equation}
which we compute directly from the evolving interior properties at the $P=1\,$Mbar surface.

Figure~\ref{fig:B_L}b displays the post-accretion evolution of $B_{2\rj}$. For the $0.3\,\mj$ planet, the magnetic field starts weak ($\sim 10^{-1}\,$G) before jumping abruptly to a few gauss at $\sim 1\,$Myr and subsequently decaying. This rapid amplification is triggered when the $P=1\,$Mbar layer transitions from a radiative to a convective state. In this low-mass regime, the ``photospheric'' and ``realistic EoS'' tracks remain closely aligned.

At higher planetary masses ($1.0\,\mj$ and $3.0\,\mj$), strong dynamo fields are established immediately after accretion. For the $1.0\,\mj$ planet, $B_{2\rj}$ starts at $\sim 100\,$G and declines gradually to $\sim 30\,$G by $10\,$Myr, exhibiting only minor sensitivity to the outer boundary condition. For the $3.0\,\mj$ planet, however, the boundary condition exerts a pronounced effect. The ``realistic EoS'' model generates an initial field of several hundred gauss, more than double the $\sim 150\,$G field of the ``photospheric'' model, and sustains a systematically stronger field throughout the $10\,$Myr evolution. This enhanced dynamo output directly reflects the hotter interior and larger convective energy flux produced by shock-driven \ce{H2} dissociation.

\section{Discussion} \label{sec:discussion}

\subsection{Coupling accretion physics and interior evolution}

Previous studies of gas giant formation and evolution have typically initialized young planets using constant-$\gamma$ analytic prescriptions \citep{Berardo2017}, broad random sampling of initial thermodynamic states \citep{knierim2025}, or idealized cold- and hot-start assumptions \citep{sur2025}. In this work, we present a framework that directly couples RHD simulations of the accretion shock, including post-shock \ce{H2} dissociation, with long-term interior evolution calculations in {\tt MESA}.

By comparing models evolved under default ``photospheric'' and our ``realistic EoS'' outer boundary conditions across three representative accretion histories (Sect.~\ref{sec:internalevolution}), we quantify how shock microphysics shapes the nascent planet. Overall, the divergence between the two boundary prescriptions grows progressively with planetary mass. While \ce{H2} dissociation begins to influence post-shock entropy around $M_{\rm final} \gtrsim 1\,\mj$, it becomes a dominant driver of the thermal structure for massive gas giants ($M_{\rm final} \gtrsim 3\,\mj$).

Our interior models incorporate several simplifying assumptions, including an inert core \citep{Berardo2017}, a homogeneous envelope composition that neglects primordial composition gradients, and the SCVH EoS \citep{saumon1995}, which overestimates the hydrogen metallization pressure. Although these approximations affect absolute interior profiles, they apply equally to both sets of simulations and therefore do not alter our differential comparison between the ``photospheric'' and ``realistic EoS'' boundary conditions. Nevertheless, because a planet's initial thermodynamic state critically governs both its subsequent internal evolution \citep{vazan2018} and the architecture of its circumplanetary environment and satellite system \citep{batygin2018,batygin2025}, incorporating self-consistent composition gradients and updated equations of state remains an important avenue for future work.

\subsection{Implications for observations and early evolution}

High-contrast direct imaging has revealed several young protoplanets still embedded within their natal protoplanetary disks, detected via either infrared thermal continuum or \ce{H\alpha} line emission. Disentangling whether the observed infrared flux is dominated by intrinsic planetary emission or ongoing accretion luminosity is crucial for inferring planetary masses and evolutionary states. As shown in Fig.~\ref{fig:B_L}, our ``realistic EoS'' boundary conditions yield intrinsic luminosities $4$--$10$ times higher than those predicted by standard cold-start models. Consequently, applying cold-start evolutionary tracks to directly imaged planets significantly underestimates their intrinsic brightness at a given mass, which in turn leads to an overestimate of the inferred planetary mass, particularly in the high-mass regime ($M_{\rm final} \gtrsim 3\,\mj$).

Conversely, \ce{H\alpha} emission is widely interpreted as a tracer of magnetospheric accretion and used to estimate mass accretion rates. Taking the well-studied protoplanet PDS~70c as an example, its observed \ce{H\alpha} luminosity of $L_{\mathrm{H\alpha}} \gtrsim 2.6 \times 10^{-7}\,\mathrm{L_\odot}$ \citep{haffert2019,close2025a,zhou2025} corresponds to a total accretion luminosity of $L_{\mathrm{acc}} \gtrsim 2.6 \times 10^{-5}\,\mathrm{L_\odot}$ under the assumption of pure magnetospheric accretion \citep{aoyama2019}. However, this value is lower than the combined near- to mid-infrared photometric luminosity measured across the $J$, $H$, $K$, $L'$, and $M'$ bands \citep{mesa2019,wang2020,wang2021,christaens2024,blakely2025}, and the true bolometric luminosity of PDS~70c must be higher still. A substantial fraction of PDS~70c's emission may therefore originate from the planet's intrinsic cooling luminosity and its circumplanetary disk \citep{chen2026b}. Accurately predicting intrinsic planetary luminosities is thus essential for interpreting multi-wavelength observations of forming planets.

Finally, the planetary magnetic field is generated by dynamo action within the deep metallic hydrogen envelope. Applying the energy-flux scaling relation of \citet{christensen2009} following \citet{Fujii2026}, our predicted field strengths (Fig.~\ref{fig:B_L}) represent theoretical upper limits, as the scaling assumes magnetic energy equipartition in the rapid-rotation regime. Even under this optimistic assumption, a Jupiter-mass planet's equatorial dipole field at $2\,\rj$ declines from $\sim 100\,$G to $\sim 30\,$G over the first $10\,$Myr. Such field strengths may be insufficient to magnetically truncate the circumplanetary disk around young Jupiter-mass planets \citep{batygin2025}, underscoring the need to explore a wider range of accretion histories with coupled evolutionary models.

\section{Conclusions} \label{Sec:conclusion}

In this work, we systematically investigated how \ce{H2} dissociation during runaway gas accretion shapes the initial thermal and magnetic states of young gas giants. Using a radiation-hydrodynamic framework \citep{Chen2022} coupled with a realistic equation of state incorporating \ce{H2} dissociation via the Saha equation, we conducted 1000 steady-state accretion shock simulations with {\tt Guangqi} \citep{chen2026} across a broad grid of planetary masses ($\mpl$), radii ($\rpl$), accretion rates ($\dot{M}$), and internal luminosities ($L_{\mathrm{int}}$). From this dataset, we trained a physics-informed neural network (PINN) surrogate model (Appendix~\ref{A:pinn_training}) to rapidly and accurately predict post-shock thermodynamic states. Coupling this surrogate model to {\tt MESA} enabled us to self-consistently co-evolve the planetary interior structure and post-shock outer boundary conditions across prescribed accretion histories.

Our main findings are summarized as follows:
\begin{enumerate}
    \item The post-shock specific entropy predicted by the PINN substantially exceeds the photospheric entropy typically adopted in cold-start models, making the planets with PINN-derived boundary conditions 4--10 times brighter than their cold-start counterparts (Fig.~\ref{fig:B_L}a). For massive protoplanets ($\mpl > 1\,\mj$) undergoing rapid accretion, partial \ce{H2} dissociation occurs naturally at the accretion shock (Fig.~\ref{fig:3mj}). Consequently, newly formed massive gas giants ($M_{\rm final} \gtrsim 3\,\mj$) attain intrinsic luminosities of $\sim 10^{-4}\,\mathrm{L_\odot}$ during their first $1\,$Myr (Sect.~\ref{sec:evolution})—roughly an order of magnitude brighter than their cold-start counterparts.

    \item Using the convective energy flux within the metallic hydrogen dynamo region ($P > 1\,$Mbar; \citealt{christensen2009}), we derived upper limits on the equatorial dipole magnetic field strength at $2\,\rj$ ($B_{2\rj}$; Sect.~\ref{sec:dynamo}) under fiducial accretion histories:
    \begin{itemize}
        \item A Saturn-mass planet ($0.3\,\mj$) generates a weak dipole field of only a few gauss, which is unlikely to truncate its circumplanetary disk (CPD) and points toward a distinct satellite formation pathway \citep{fujii2020,Fujii2026}.
        \item A Jupiter-mass planet ($1.0\,\mj$) begins with $B_{2\rj} \sim 100\,$G, decaying to $\sim 30\,$G over $10\,$Myr. Because these values likely represent theoretical upper limits, magnetospheric accretion from the CPD is expected to operate only once the disk surface density drops to low values \citep[e.g., $<0.01\,\mathrm{g\,cm^{-2}}$;][]{fujii2014}, at which point the low-mass CPD cannot efficiently spin down the planet via magnetic braking \citep{batygin2018}.
        \item A super-Jupiter ($3.0\,\mj$) remains intrinsically luminous ($10^{-5}$--$10^{-4}\,\mathrm{L_\odot}$) over its first $10\,$Myr and begins with a strong dipole field of $B_{2\rj} \sim 500\,$G, decaying to $\sim 150\,$G over $10\,$Myr. Such a field is capable, in principle, of driving magnetospheric accretion \citep{taylor2024}, though the resulting accretion rate depends sensitively on the CPD ionization fraction and warrants further study.
    \end{itemize}
\end{enumerate}

The elevated initial luminosities of massive gas giants significantly enhance their detectability via high-contrast direct imaging. By retaining substantial shock-generated heat, young giants remain bright at near- and mid-infrared wavelengths for hundreds of millions of years \citep{marois2008}, creating strong synergies with current and upcoming direct-imaging surveys \citep{chilcote2020,desidera2021,zhu2026}. Expanding the demographic sample of directly imaged young exoplanets will place stringent observational constraints on post-formation entropy and directly test the accretion shock physics modeled here.

\begin{acknowledgements}
This work is financially supported by the Tsinghua University Dushi Program and National Natural Science Foundation of China under grant No. 12342501. The Center for High Performance Computing at Tsinghua University provided the computational resources. We also acknowledge the Chinese Center for Advanced Science and Technology for hosting the Protoplanetary Disk and Planet Formation Summer School in 2024, which inspired this work.
\end{acknowledgements}

\bibliographystyle{aa} 
\bibliography{ref} 

\begin{appendix}
\section{Physics-informed loss function derivation} \label{AA}

The specific entropy per unit mass of an ideal gas mixture is written as
\begin{equation}
    s = \frac{k_{\rm B}}{\rho}\sum_in_i(1+\frac{\dd \ln{z_i}}{\dd \ln{T}}-\ln{\frac{n_i}{z_i}}) \, ,
    \label{s}
\end{equation}
where $k_{\rm B}$, $T$, and $\rho$ denote the Boltzmann constant, gas temperature, and mass density, respectively, while $n_i$ and $z_i$ represent the number density and partition function of species $i \in \{\mathrm{H}, \mathrm{H_2}, \mathrm{He}\}$.

Neglecting nuclear spin, electronic excitation, and molecular rotational--vibrational terms, the translational partition functions reduce to
\begin{equation}
    z_{\rm H} = 
    \left(\frac{2m_{\rm H}\pi k_{\rm B}T}{h^2}\right)^{\frac{3}{2}}
    \exp{-\frac{\phi_{\rm dis}}{2k_{\rm B} T}} \, ,
    \label{zh}
\end{equation}
\begin{equation}
    z_{\rm H_2} = \left(\frac{2m_{\rm H_2}\pi k_{\rm B}T}{h^2}\right)^{\frac{3}{2}} \, ,
    \label{zh2}
\end{equation}
\begin{equation}
    z_{\rm He} = \left(\frac{2m_{\rm He}\pi k_{\rm B}T}{h^2}\right)^{\frac{3}{2}} \, ,
    \label{zhe}
\end{equation}
where $h$ is the Planck constant and $\phi_{\rm dis}$ is the dissociation energy per \ce{H2} molecule.

Considering only the dissociation reaction $\mathrm{H_2} \rightarrow 2\mathrm{H}$, the number densities $n_{\rm H_2}$ and $n_{\rm H}$ are governed by hydrogen atom conservation and the Saha equation:
\begin{equation}
    2n_{\rm H_2}+n_{\rm H} = \frac{X\rho}{m_{\rm H}} \, , 
    \label{Hconservation}
\end{equation}
\begin{equation}
    \frac{n_{\rm H}^2}{n_{\rm H_2}} = \frac{z_{\rm H}^2}{z_{\rm H_2}} \, ,
    \label{saha}
\end{equation}
where $X=0.74$ is the hydrogen mass fraction adopted in our simulations and $m_{\rm H}$ is the atomic hydrogen mass. Substituting Eqs.~(\ref{zh})--(\ref{saha}) into Eq.~(\ref{s}) simplifies the specific entropy to
\begin{equation}
\begin{aligned}
s = \frac{k_{\rm B}}{\rho}\Bigg[
    &\frac{5}{2}\left(n_{\rm H_2}+n_{\rm H}+n_{\rm He}\right)\\
    &{}+ \frac{\phi_{\rm dis}}{2k_{\rm B} T} n_{\rm H} - \frac{X\rho}{m_{\rm H}} \ln{\frac{n_{\rm H}}{z_{\rm H}}}
- n_{\rm He} \ln{\frac{n_{\rm He}}{z_{\rm He}}}
\Bigg] \, .
\end{aligned}
\end{equation}
Combining Eqs.~(\ref{Hconservation}) and~(\ref{saha}) yields the equilibrium number densities for each species:
\begin{equation}
    n_{\rm H} = \frac{-1+\sqrt{\Delta}}{\frac{2^{3.5}}{z_{\rm H}}\exp{\frac{\phi_{\rm dis}}{2k_{\rm B} T}}}\, , \qquad \Delta = 1+\frac{2^{4.5}X\rho}{z_{\rm H}m_{\rm H}}\exp{\frac{\phi_{\rm dis}}{2k_{\rm B} T}} \, ,
\end{equation}
\begin{equation}
    n_{\rm H_2} = \frac{X\rho/m_{\rm H} - n_{\rm H}}{2} \, ,
\end{equation}
\begin{equation}
    n_{\rm He} = \frac{(1-X)\rho}{4m_{\rm H}} \, .
\end{equation}

To improve numerical stability and prevent vanishing or exploding gradients during optimization, all input features are standardized prior to training. Because standardization rescales the physical quantities, the original dimensional constants cannot be applied directly. We therefore introduce three global trainable coefficients—$a_1$, $a_2$, and $a_3$—that are updated automatically alongside the network weights:
\begin{equation}
    a_1 \equiv \left(\frac{2\pi k_{\rm B}}{h^2}\right)^{\frac{3}{2}}\, , \qquad
    a_2 \equiv \frac{\phi_{\rm dis}}{2k_{\rm B}} \, , \qquad
    a_3 \equiv m_{\rm H} \,.
    \label{eqn:abc}
\end{equation}
In terms of $a_1$, $a_2$, and $a_3$, the partition functions become
\begin{equation}
    z_{\rm H} = a_1\left(a_3T\right)^{\frac{3}{2}}\exp{-\frac{a_2}{T}} \, , \quad
    z_{\rm H_2} = a_1\left(2a_3T\right)^{\frac{3}{2}} \, , \quad
    z_{\rm He} = a_1\left(4a_3T\right)^{\frac{3}{2}} \, ,
\end{equation}
and the species number densities are given by
\begin{equation}
    n_{\rm H} = \frac{2X\rho/a_3}{1+\sqrt{\Delta}}\,, \qquad
    \Delta = 1+\frac{2^{4.5}X}{z_{\rm H}a_3}\rho \exp{\frac{a_2}{T}}
\end{equation}
\begin{equation}
    n_{\rm H_2} = \frac{X\rho/a_3 - n_{\rm H}}{2} \, , \qquad
    n_{\rm He} = \frac{(1-X)}{4a_3}\rho
\end{equation}
Omitting the multiplicative constant $k_{\rm B}$ during PINN training, the normalized specific entropy and pressure evaluated in the physics-informed loss terms are
\begin{equation}
\begin{aligned}
s = \frac{1}{\rho}\Bigg[
    &2.5 \left( n_{\rm H}+n_{\rm H_2}+n_{\rm He} \right)\\
    &{} + \frac{a_2}{T} n_{\rm H} - \frac{X\rho}{a_3} \cdot \ln \left( \frac{n_{\rm H}}{z_{\rm H}} \right) -n_{\rm He}\ln{\left(\frac{n_{\rm He}}{z_{\rm He}} \right)}
    \Bigg] \, ,
    \label{loss-s}
\end{aligned}
\end{equation}
and
\begin{equation}
    P = \left(n_{\rm H}+n_{\rm H_2}+n_{\rm He}\right)T \ .
    \label{loss-p}
\end{equation}

\section{Comparison between MLP and PINN} \label{A:pinn_training}

\begin{figure*}[h!]
    \centering
    \includegraphics[width=1\linewidth]{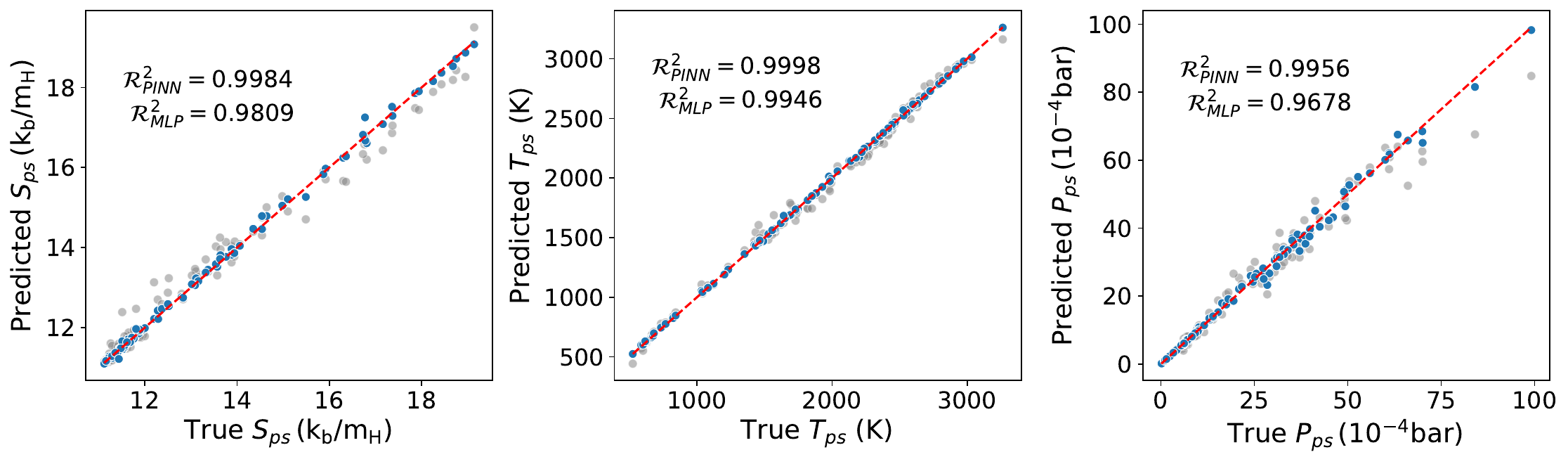}
    \caption{Predicted versus simulated post-shock quantities ($s_{\mathrm{ps}}$, $T_{\rm ps}$, and $P_{\rm ps}$) on the test set, with the corresponding $\mathcal{R}^2$ values indicated in each panel. Blue and grey markers denote predictions from the PINN and baseline MLP models, respectively.}
    \label{fig:pinnSTP}
\end{figure*}

\begin{figure*}[h!]
    \centering
    \includegraphics[width=1\linewidth]{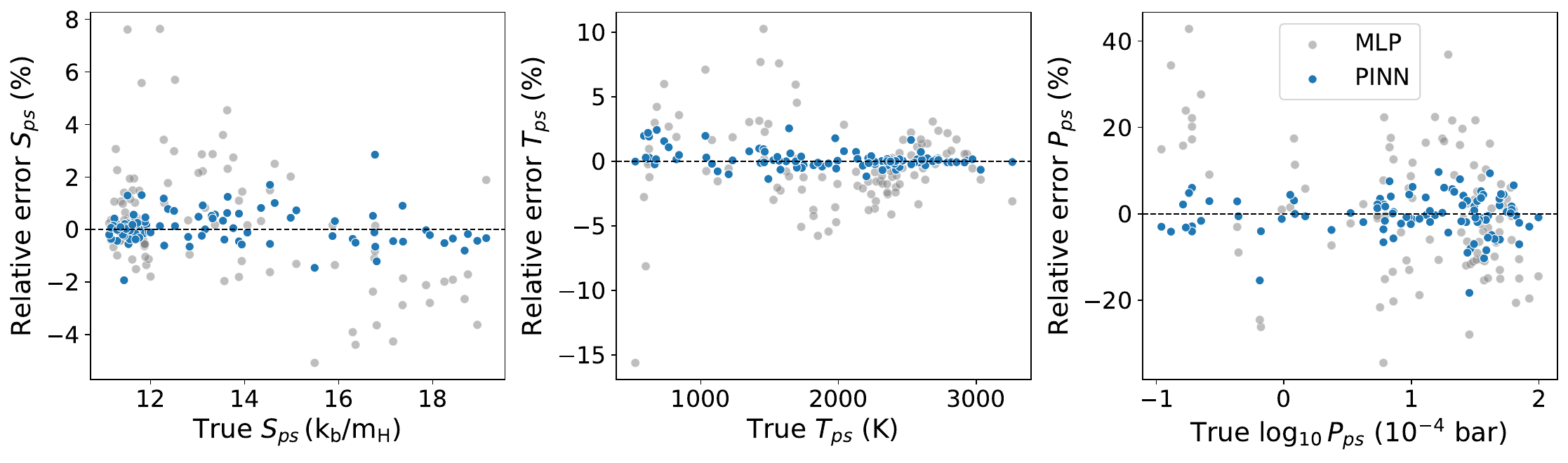}
    \caption{Relative prediction errors for (a) post-shock specific entropy, (b) post-shock temperature, and (c) post-shock pressure across the test set. Black dashed lines denote zero residual.}
    \label{fig:resplotSTP}
\end{figure*}

We evaluate model accuracy on an independent test set of $N=100$ additional {\tt Guangqi} simulations (drawn separately from the 1000-model training survey) using the coefficient of determination ($\mathcal{R}^2$) and the mean absolute error (MAE):
\begin{equation}
    \mathcal{R}^2 = 1-\frac{\sum_{i=1}^{N} \left(x_0^{(i)}-x^{(i)}\right)^2}{\sum_{i=1}^{N}\left(x_0^{(i)}-\overline{x}_0\right)^2} \, ,
    \label{eqn:r2}
\end{equation}
\begin{equation}
    \mathrm{MAE} = \frac{1}{N} \sum_{i=1}^{N} \left|x^{(i)}-x_0^{(i)}\right| \, ,
    \label{eqn:mae}
\end{equation}
where $x \in \{s_{\mathrm{ps}}, T_{\rm ps}, P_{\rm ps}\}$, $x^{(i)}$ and $x_0^{(i)}$ denote the predicted and simulated values for sample $i$, and $\overline{x}_0$ is the sample mean of the simulated target.

Figure~\ref{fig:pinnSTP} compares the test-set predictions of the trained PINN against those of a baseline MLP across all three simultaneously predicted post-shock variables, while Fig.~\ref{fig:resplotSTP} shows the corresponding relative residuals. The PINN achieves close agreement with the {\tt Guangqi} simulations, exceeding $\mathcal{R}^2 > 0.99$ across all three variables and reducing the MAE from $0.25\,k_{\rm B}\,m_{\rm H}^{-1}$, $77.2\,$K, and $2.38 \times 10^{-4}\,$bar (MLP) to $0.10\,k_{\rm B}\,m_{\rm H}^{-1}$, $23.9\,$K, and $1.23 \times 10^{-4}\,$bar (PINN), respectively. As shown in Fig.~\ref{fig:resplotSTP}, the baseline MLP struggles at the extremes of the parameter space: it systematically overestimates small values and underestimates large values of $s_{\mathrm{ps}}$ and $P_{\rm ps}$, and exhibits large errors in $T_{\rm ps}$ below $2000\,$K. Incorporating physical constraints in the PINN effectively eliminates these systematic biases.

Training the PINN converges within $20$--$30\,$min, with both training and validation losses decreasing rapidly before plateauing smoothly, validating our early-stopping criterion. Once trained, the network predicts post-shock thermodynamic states in $<10^{-4}\,$s per query—compared to approximately $20\,$CPU hours required for a single full {\tt Guangqi} simulation.

Several methodological aspects warrant further exploration. Whereas generalization in standard neural networks is typically assessed by comparing training and validation losses, rigorously evaluating out-of-domain generalization in PINNs is more complex \citep{bonfanti2024} and lies beyond the scope of this work. Confirming reliable extrapolation would reduce dependence on large parameter surveys, allowing the PINN to predict post-shock states beyond the parameter boundaries sampled by {\tt Guangqi}. Furthermore, because our architecture maps multiple inputs to multiple outputs while jointly optimizing three global physical coefficients, our 1000-model dataset represents a relatively modest sample size. Although PINNs are well suited to data-scarce regimes \citep{bonfanti2024}, expanding the training survey to $\sim 10^{4}$ {\tt Guangqi} simulations in future work will further improve predictive robustness across the full parameter space.

\end{appendix}

\end{document}